\documentclass[journal]{IEEEtran}

\usepackage{amsmath}
\usepackage{multirow}
\usepackage{subfigure}
\usepackage{graphicx}
\usepackage{algorithm}
\usepackage{algorithmicx}
\usepackage[noend]{algpseudocode}
\usepackage{color}
\usepackage{url}
\usepackage{tikz}
\usepackage{amsfonts}
\usepackage{array}
\usepackage{gensymb}
\usepackage{comment}
\newenvironment{conditions}[1][where:]
  {#1 \begin{tabular}[t]{>{$}l<{$} @{${}={}$} l}}
  {\end{tabular}\\[\belowdisplayskip]}

\newcounter{reviewercount}
\newcounter{commentcount}

\newcommand{\aeditor}%
  {\bigskip\noindent {\bf COMMENTS OF THE ASSOCIATE EDITOR}%
  \setcounter{commentcount}{0}\par 
}

\graphicspath{{FIGS/}}

\begin{document}

\setcounter{page}{1}
\twocolumn

\title{PROBE3.0: A Systematic Framework for Design-Technology 
Pathfinding with Improved Design Enablement}

\author{
Suhyeong~Choi,~\IEEEmembership{Graduate Student Member,~IEEE,}
Jinwook~Jung,~\IEEEmembership{Member,~IEEE,}\\
Andrew~B.~Kahng,~\IEEEmembership{Fellow,~IEEE,}
Minsoo~Kim,~\IEEEmembership{Member,~IEEE,}
Chul-Hong~Park,~\IEEEmembership{Member,~IEEE,}
Bodhisatta~Pramanik,~\IEEEmembership{Graduate Student Member,~IEEE,} and
Dooseok~Yoon,~\IEEEmembership{Graduate Student Member,~IEEE}
}

\maketitle

\begin{abstract}
We propose a systematic framework to conduct 
design-technology pathfinding for PPAC
in advanced nodes. 
Our goal is to provide configurable, scalable generation of process 
design kit (PDK) and standard-cell library, spanning key scaling boosters
(backside PDN and buried power rail), to explore PPAC across given 
technology and design parameters.
We build on \cite{ChengKKK22}, which addressed only area and cost (AC),
to include power and performance (PP) evaluations through automated
generation of full design enablements.
We also improve the use of artificial designs in the PPAC assessment of
technology and design configurations. 
We generate more realistic artificial designs by applying a
machine learning-based parameter tuning flow to ~\cite{KimLMK22}.
We further employ {\em clustering-based cell width-regularized 
placements} at the core of routability assessment, enabling more
realistic placement utilization and improved experimental efficiency.
We demonstrate PPAC evaluation across scaling boosters and 
artificial designs in a predictive technology node.
\end{abstract}

\vspace{-0.5cm}
\section{Introduction}
\label{sec:intro}

Due to the slowdown of dimension scaling relative to the trend of 
the traditional Moore's Law, scaling boosters, such as backside power delivery networks (BSPDN),
buried power rails (BPR), are introduced at advanced technology nodes.
Since scaling boosters play an important role to optimize
power, performance, area and cost (PPAC) of advanced technologies,
accurate and fast evaluations and predictions of PPAC
are critical at an early stage of technology development.
Also, use of scaling boosters makes evaluations and predictions of technology
more difficult since they introduce a large number of knobs to contribute PPAC 
improvement.

\begin{figure}[t]
\center
\includegraphics[width=0.95\columnwidth]{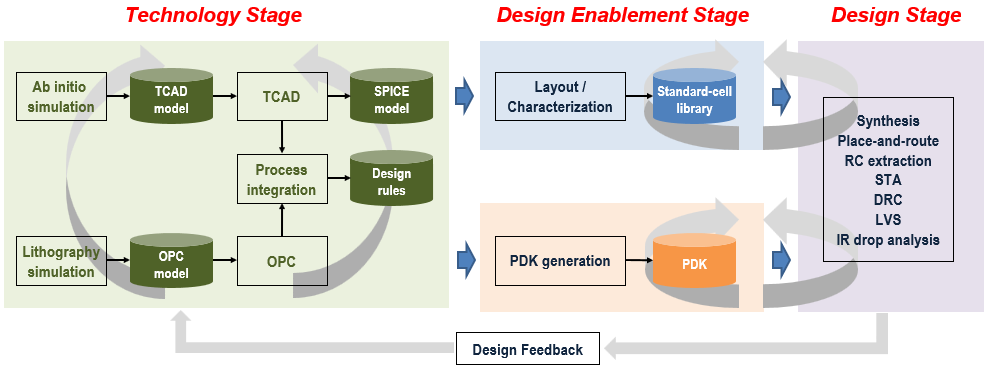}
\caption{The DTCO process consists of  
three main stages: {\em Technology}, {\em Design Enablement},
and {\em Design}. 
Figure is redrawn from~\cite{SynopsysDTCO}.} 
\vspace{-0.3cm}
\label{fig:dtco}
\end{figure}

% DTCO is important but slow and complicated due to scaling booster
Design-Technology Co-Optimization (DTCO) is now a well-known key 
element to develop advanced technology nodes and designs in the 
modern VLSI chip design.
Today's DTCO spans assessment and co-optimization across almost
all components of semiconductor technology and design enablement.
As described in Figure~\ref{fig:dtco}, the DTCO process consists of three stages: 
{\em Technology}, {\em Design Enablement}, and {\em Design}.
First, the technology stage includes modeling and simulation 
methodologies for process and device technology. 
Second, the design enablement stage includes creation of required 
process design kits (PDK) for the ensuing design stage; 
these include device models, standard-cell libraries, 
routing technology files and interconnect parasitic (RC) models.
Last, the design stage includes logic synthesis and 
place-and-route (P\&R) based on the generated PDKs from the design 
enablement stage.

To evaluate and predict technology and design at advanced nodes, 
all three stages must be correctly performed, and PDKs must be 
generated from technology and design enablement stages.
However, the DTCO process is not simple:
feedback from design stage to technology stage takes 
weeks to months of turnaround time, along with immense engineering efforts.
Also, based on the design feedback, additional PDKs may need 
to be generated at the design enablement stage, which requires additional
weeks to months.
In order to reduce the turnaround time and maximize the benefit 
of the DTCO process, a fast and accurate DTCO methodology is needed
to assess PPAC with reasonable turnaround time, and to more precisely guide
multi-million dollar decisions at an early stage of technology development. 

\noindent
{\bf Contributions of Our Work.}
Compared to the previous works PROBE1.0 \cite{KahngKLL18} and PROBE2.0 \cite{ChengKKK22}, 
our new framework provides three main technical achievements.

\noindent
(1) \textbf{We establish the first comprehensive end-to-end design and technology pathfinding framework.}
    \cite{ChengKKK22}\cite{KahngKLL18} focus on area and cost 
    without considering power and performance. 
    Thus, there is a significant discrepancy between \cite{ChengKKK22}\cite{KahngKLL18} 
    and the actual DTCO process in the industry. 
    In this work, we propose a more complete and systematic 
    PROBE3.0 framework, which incorporates power and performance 
    aspects for design-technology pathfinding 
    at an early stage of technology development. PROBE3.0 enables fast and 
    accurate PPAC evaluations by generating configurable PDKs, 
    including standard-cell libraries.

\noindent
(2) \textbf{We improve our designs for PPAC explorations.}
    Design is a critical factor for PPAC explorations, and artificially 
    generated designs enable us to explore a wider solution space.
    We leverage \cite{KimLMK22} to generate artificial designs.
    To create more realistic artificial designs, we develop a machine 
    learning (ML)-based parameter tuning flow built on \cite{KimLMK22} 
    to find the best input parameters  for generating such designs.
    Section~\ref{sec:ang} details our artificial design generation flow. 
    Further, {\em cell width-regularization} is employed in 
    \cite{ChengKKK22}\cite{KahngKLL18} to prevent illegal placements when 
    swapping neighboring cells to assess the routability metric, $K_{th}$.
    We propose a {\em clustering-based cell width-regularization} 
    to achieve more realistic utilization (and faster routability assessment)
    as described in Section~\ref{sec:enhancedcw}.

\noindent
(3) \textbf{We demonstrate the PPAC exploration of scaling boosters.}
    We incorporate scaling boosters (BSPDN and BPR) to support 
    P\&R and IR drop analysis flows within the framework, as detailed in 
    Section~\ref{sec:powerdelivery}. 
    Our results show that incorporating BSPDN and BPR leads to a reduction in power consumption by up to 8\% and area by up to 24\% based on 
    our predictive 3nm technology.
    The area reduction results are consistent with those reported in 
    previous industry works~\cite{HossenCVB20}\cite{Ryckaert19}\cite{BSBPRAM21}\cite{BSBPRarticle22},
    which have demonstrated area reductions of 25\% to 30\% through the 
    use of BSPDN and BPR techniques.

Due to limited access to advanced technology for academic research,
we build our predictive 3nm technology, named {\em the PROBE3.0 technology}.
To calibrate the technology, we refer to the International Roadmap for 
Devices and Structures (IRDS)~\cite{IRDS}, open-sourced PDKs, and other  
publications~\cite{BhanushaliD15}\cite{ClarkVSG16}\cite{Sadangi21}\cite{FreePDK3}. 
We open-source our work, including process design kits (PDKs), 
standard-cell libraries, and scripts for P\&R and IR drop analysis;
this is available 
in our GitHub repository~\cite{PROBE3.0}. 
In Section~\ref{sec:probe3}, we provide details on the automated PDKs and library generation flows,
while in Section~\ref{sec:expt}, we present three experiments to demonstrate the effectiveness of the PROBE3.0 framework for PPAC pathfinding.

\begin{figure}[t]
\center
\includegraphics[width=0.75\columnwidth]{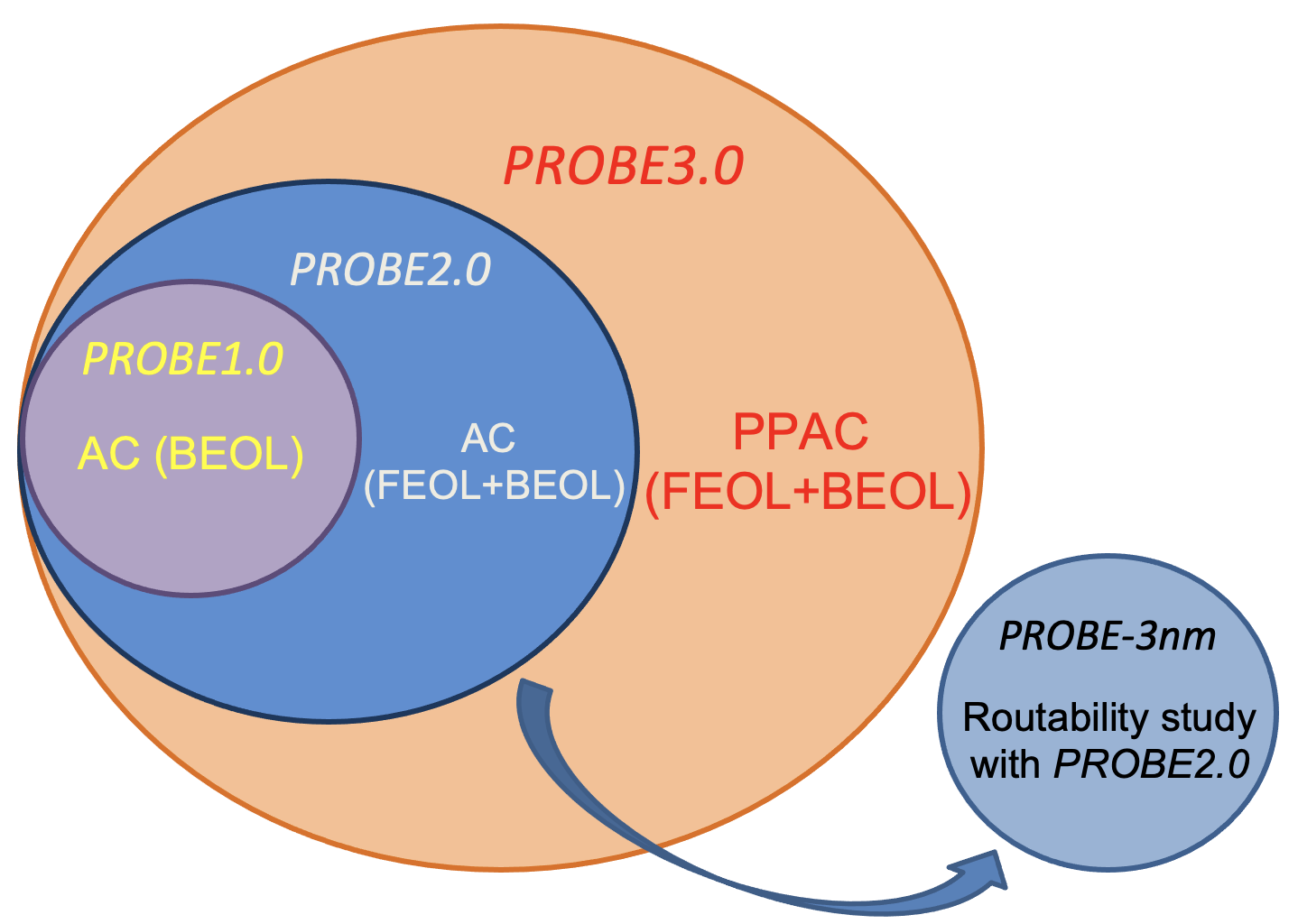}
\caption{Scope of PROBE-related works. {\em PROBE1.0}~\cite{KahngKLL18} 
and {\em PROBE2.0}~\cite{ChengKKK22}
address AC given BEOL and FEOL/BEOL, respectively. 
{\em PROBE-3nm}~\cite{ChidambaramKKN21} studies routability (AC) with sub-3nm 
technology configurations. PROBE3.0 provides true full-stack PPAC 
pathfinding with automatic generation of EDA tool enablements.}
\vspace{-0.3cm}
\label{fig:history}
\end{figure}

\vspace{-0.3cm}
\section{Related Work}
\label{sec:relatedwork}
\vspace{-0.1cm}

In this section, we divide the relevant previous works into the three categories
of (i) advanced-technology research PDKs, (ii) design-technology 
co-optimization and (iii) scaling boosters, along with
(iv) ``PROBE'' frameworks.

\noindent 
{\bf Advanced Technology Research PDKs.}
PDKs of advanced node technologies are highly confidential. 
Academic research can be blocked by limited access 
to relevant information.
To unblock academic research, predictive advanced-node PDKs have been 
published. ASAP7~\cite{ClarkVSG16} is a predictive PDK for 7nm 
FinFET technology that includes standard cells which support 
commercial logic synthesis and P\&R.
FreePDK3~\cite{Sadangi21}\cite{FreePDK3} and
FreePDK15~\cite{BhanushaliD15} 
are open-source PDKs for 3nm and 15nm technology.
\cite{KimJWY22} proposes a 3nm predictive 
technology called NS3K with nanosheet FETs (NSFET). 
The authors of~\cite{KimJWY22} also
create 5nm FinFET and 3nm NSFET libraries to compare power, performance and area.

\noindent
{\bf Design-Technology Co-Optimization.}
Previous DTCO works
evaluate block-level PPAC and optimize design and
technology simultaneously.
\cite{Song16} proposes UTOPIA to evaluate block-level PPAC with thermally
limited performance, and to optimize device and technology parameters.
\cite{LiebmannCCC20} proposes a fast pathfinding 
DTCO flow for FinFET and complementary FET (CFET).
\cite{ChanemougameSGB20} also proposes a fast and agile
technology pathfinding platform with compact device
models to accelerate the DTCO process.
\cite{KahngKKX21} describes power delivery network pathfinding for
3D IC technology to study tradeoffs between IR drop and routability. 
\cite{ChengHHL21} uses ML to predict sensitivities to changes for DTCO.

\noindent
{\bf Scaling Boosters.}
As described in Section~\ref{sec:intro}, 
scaling boosters are used in advanced nodes to 
maximize benefit of new technology. 
BSPDN and BPR are among the most promising scaling 
boosters in sub-5nm nodes. 
\cite{PrasadNDZ19} carries out a CPU implementation
with BSPDN and BPR in their 3nm technology, 
demonstrating a reduction of up to 7X in worst IR drop. 
Similarly, \cite{Ryckaert19} investigates BSPDN and BPR at sub-3nm nodes
and finds that they can lead to a 30\% reduction in area 
based on IR drop mitigation.
\cite{ChavaSJG19} also explores the impact of BSPDN and BPR 
on design, concluding that their use can lead to a 
43\% reduction in area with 4X less IR drop.
\cite{HossenCVB20} studies BSPDN configurations with 
$\mu$TSVs, and observes 25\% to 30\% reduction 
in area using BSPDN and BPR. Additionally, \cite{Sisto21} 
investigates BSPDN with nTSVs and $\mu$TSVs and finds that the 
average IR drop with BSPDN improves by 69\% compared to traditional 
frontside PDN (FSPDN). 
Finally, \cite{TejaPDZ22} conducts holistic evaluations for BSPDN and BPR, 
demonstrating that FSPDN with BPR achieves a 25\% lower on-chip 
IR drop, while BSPDN with BPR achieves an 85\% lower on-chip IR drop 
with iso-performance and iso-area.
In contrast to these previous DTCO works, here we propose a highly 
{\em configurable} framework that enables more efficient investigation of 
scaling boosters in advanced nodes.

\noindent
{\bf ``PROBE'' Frameworks.}
Prior ``PROBE'' \cite{ChengKKK22}\cite{KahngKLL18} works 
propose systematic frameworks for assessing routability with 
different FEOL and BEOL configurations. Specifically, \cite{KahngKLL18} begins
with an easily-routable placement and increases the routing difficulty 
by random neighbor-swaps until the routing fails with greater than a threshold 
number of design rule violations (DRCs). The normalized number of swaps
at which routing failure occurs, denoted by $K_{th}$, is a metric used 
to measure the inherent routability of the given parameters. 
On the other hand, \cite{ChengKKK22} introduces an automatic standard-cell layout generation
using satisfiability modulo theory (SMT) to support explorations of 
both FEOL and BEOL configurations. The authors of \cite{ChengKKK22}
also employ machine learning (ML)-based 
$K_{th}$ prediction to expedite the DTCO pathfinding process. 
Additionally, \cite{ChidambaramKKN21} employs PROBE2.0 in a 
routability study with sub-3nm technology configurations.

\vspace{-0.3cm}
\section{Standard-Cell Library and PDK Generation}
\label{sec:probe3}
\vspace{-0.1cm}

Expediting the DTCO process requires automation of the standard-cell library
and PDK generation flows. Therefore, the PROBE2.0 framework~\cite{ChengKKK22}
introduces standard-cell layout and PDK generation flows and 
utilizes them for routability assessments. In this work, we extend 
the PROBE2.0 framework to include proper electrical models of standard-cell libraries
and interconnect layers for design-technology pathfinding. Additionally, we enhance
the PDK generation flow to support advanced nodes. While the PROBE2.0 framework 
solely focuses on the physical layout of standard cells, the PROBE3.0 framework 
enables true full-stack PPAC pathfinding through automated, configurable 
standard-cell and PDK generation flows for advanced nodes. 
To demonstrate use of PROBE3.0 for advanced-node PPAC pathfinding,
we use a technology that incorporates cutting-edge (3nm FinFET) 
technology predictions based on the works of~\cite{ClarkVSG16}\cite{IRDS}. 

\begin{figure}[t]
\center
\includegraphics[width=0.97\columnwidth]{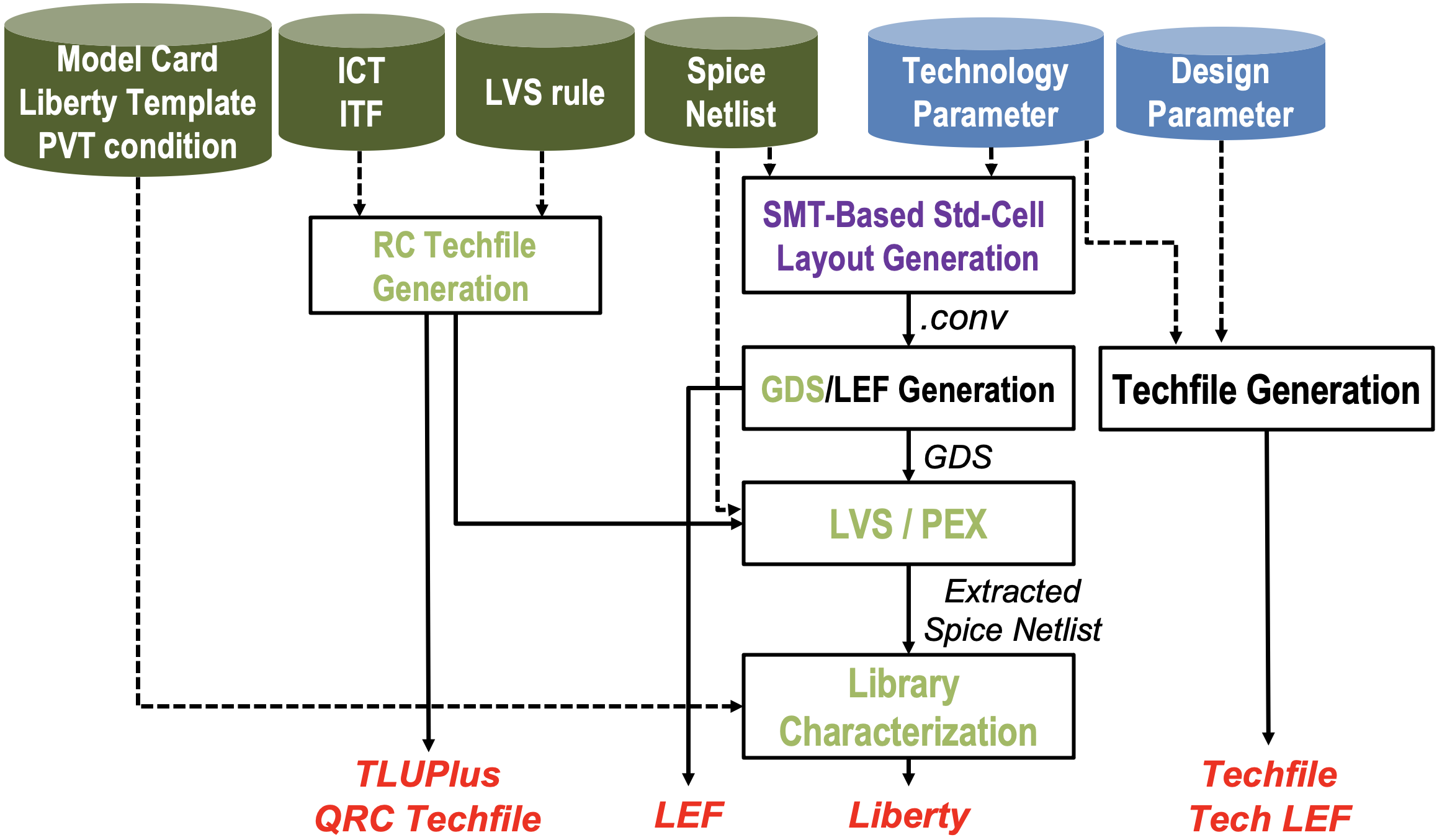}
\vspace{-0.2cm}
\caption{Automatic standard-cell library and PDK generation ({\em Design Enablements}) in the PROBE3.0 framework. In addition to technology 
and design parameters in the PROBE3.0 framework, other technology-related 
inputs are required: (i) Device model cards, (ii) Liberty templates, 
(iii) Process/Voltage/Temperature (PVT) conditions, (iv) Interconnect technology 
files (ICT or ITF formats), (v) LVS rules, and (vi) SPICE netlists.} 
\vspace{-0.5cm}
\label{fig:de_overall}
\end{figure}

\vspace{-0.4cm}
\subsection{Overall flow}
\label{subsec:overallflow}
\vspace{-0.1cm}

Figure~\ref{fig:de_overall} describes our overall flow of 
standard-cell and PDK generation.
Technology and design parameters 
are defined as input parameters for the flow.
Beyond these input parameters, there are additional inputs required
to generate standard-cell libraries and PDKs, as follows: 
(i) SPICE model cards, (ii) Liberty template and PVT conditions, 
(iii) Interconnect technology files (ICT/ITF), (iv) LVS rule deck, 
and (v) SPICE netlists.
Given the inputs, 
our SMT-based standard-cell 
layout generation and GDS/LEF generation are executed sequentially.
Generation of timing and power models (Liberty) requires
additional steps including LVS, parasitic extraction and library 
characterization flow.
Aside from the standard-cell library generation, we also 
generate interconnect models from ICT/ITF, and P\&R routing technology files 
from technology and design parameters.
The PDK elements that we generate feed seamlessly
into commercial logic synthesis and P\&R tools.
Further, to the best of our knowledge, ours is the 
first-ever work that is able to disseminate
all associated EDA tool scripts for research purposes.

\vspace{-0.5cm}
\subsection{PROBE3.0 Technology}
\label{subsec:techdef}
\vspace{-0.1cm}

We build our own predictive 
3nm technology node, called {\em the PROBE3.0 technology}.
Based on~\cite{ClarkVSG16}, we define our FEOL and BEOL layers 
as described in Table~\ref{tab:layer}.
We assume that all BEOL layers are
unidirectional routing layers.  Hence, we first change M1 
to a unidirectional routing layer with vertical preferred direction,
since the work of~\cite{ClarkVSG16} has a bidirectional M1
routing layer. 
We add an M0 layer with horizontal 
preferred direction below the modified M1 layer,
and we add contact layers V0 and CA which respectively
connect between M1 and M0, and between gate/source-drain and M0.

\begin{table}[t]
\vspace{-0.2cm}
\caption{Layer definition in the PROBE3.0 technology.}
\vspace{-0.4cm}
\scriptsize
\label{tab:layer}
\begin{center}
\begin{tabular}{|c|c|l|}
\hline
{\bf Layer} & {\bf Name}  & {\bf Description} \\\hline
\multirow{12}{*}{FEOL} & WELL & N-Well\\\cline{2-3}
 & FIN & Fin\\\cline{2-3}
 & GATE & Poly (gate)\\\cline{2-3}
 & GCUT & Gate cut\\\cline{2-3}
 & ACTIVE & Active area for fin definition\\\cline{2-3}
 & NSELECT & N-implant\\\cline{2-3}
 & PSELECT & P-implant\\\cline{2-3}
 & CA & Contact (via) between LIG/LISD and M0\\\cline{2-3}
 & LIG & Gate interconnect layer\\\cline{2-3}
 & LISD & Source-drain interconnect layer\\\cline{2-3}
 & SDT & Source-drain trench (ACTIVE to LIG/LISD)\\\cline{2-3}
 & BOUNDARY & Boundary layer for P\&R\\\hline
\multirow{2}{*}{BEOL} & M0-M13 & Metal layers\\\cline{2-3}
& V0-V12 & Via layers\\\hline
\end{tabular}
\end{center}
\vspace{-0.4cm}
\end{table}

% ITF/ICT/MIPT FEOL BEOL R thickness
Also, electrical features of technologies are critical to explore ``PP'' aspects.
Therefore, parasitic extractions of standard cells and BEOL metal stacks are 
important steps.
To extract parasitic elements,
interconnect technology files are required to use commercial RC extraction, P\&R 
and IR drop analysis tools.
In this work, we use commercial tools~\cite{QRC}\cite{Calibre}\cite{StarRC} 
for extractions, and each tool has its own technology file 
format.\footnote{The file formats for each tool are unique. The MIPT file 
format is for {\em Siemens Calibre}~\cite{Calibre} for extraction, and 
is converted to an RC rule file for standard-cell layout extractions.
On the other hand, the ICT and ITF file formats are for 
{\em Cadence} and {\em Synopsys} extraction tools, respectively. 
We convert ICT to QRC techfile, and ITF to TLUPlus file, 
to enable P\&R tools and IR drop analysis.}
Interconnect technology files include layer structures of technology and 
electrical parameters, such as thickness, width, resistivity, dielectric 
constant and via resistance.
We refer to the values of physical features in the 3nm FinFET 
technology of~\cite{IRDS}, such as fin pitch, fin width, gate pitch, 
gate width, metal pitch and aspect ratio.
We also refer to \cite{IRDS} for the values for electrical parameters
such as via resistance and dielectric constant.
Table~\ref{tab:probe3tech} describes key features of the PROBE3.0 technology.

\begin{table}[t]
\vspace{-0.2cm}
\caption{Key features of the PROBE3.0 technology.}
\vspace{-0.4cm}
\scriptsize
\label{tab:probe3tech}
\begin{center}
\begin{tabular}{|c|c|c|}
\hline
{\bf Layer} & {\bf Feature} & {\bf Value} \\\hline
\multirow{6}{*}{FEOL} & Fin Pitch & 24 nm \\\cline{2-3}
& Fin Width & 6 nm \\\cline{2-3}
& Gate Pitch (CPP) & 45 nm\\\cline{2-3}
& Gate Width & 16 nm \\\cline{2-3}
& Standard-Cell Height & 100 / 120 / 144 nm\\\cline{2-3}
& Dielectric constant & 3.9\\\hline
\multirow{10}{*}{BEOL} & Aspect Ratio (width/thickness) & 1.5 \\\cline{2-3}
& Power/Ground Pin Width (M0) & 36 nm \\\cline{2-3}
& M0/M2/M3 pitch & 24 nm \\\cline{2-3}
& M1 pitch & 30 nm \\\cline{2-3}
& M4-M11 pitch & 64 nm \\\cline{2-3}
& M12-M13 / BM1-BM2 pitch & 720 nm \\\cline{2-3}
& V0-V3 via resistance & 50 ohm/via \\\cline{2-3}
& V4-V11 via resistance & 5 ohm/via \\\cline{2-3}
& V12 via resistance & 0.06294 ohm/via \\\cline{2-3}
& Dielectric constant & 2.5-3 \\\hline
\end{tabular}
\end{center}
\vspace{-0.5cm}
\end{table}

\vspace{-0.3cm}
\subsection{Improved Standard-Cell Library Generation}
\label{subsec:stdgen}
\vspace{-0.1cm}

We generate standard-cell libraries via several steps
illustrated in Figure~\ref{fig:de_overall}: 
(i) SMT-based standard-cell layout generation, 
(ii) generation of GDS and LEF files, 
(iii) LVS and PEX flow, and 
(iv) library characterization flow.

\noindent
{\bf SMT-Based Standard-Cell Layout Generation.}
In recent technology nodes, standard-cell architectures use
a variety of pitch values for different layers in order
to optimize power, performance, area and cost (PPAC).
To accommodate this, PROBE3.0 improves the SMT-based layout generation 
used in PROBE2.0 to support non-unit gear ratios for M1 pitch (M1P) and 
contacted poly pitch (CPP).

\begin{figure}[t]
\center
\includegraphics[width=0.97\columnwidth]{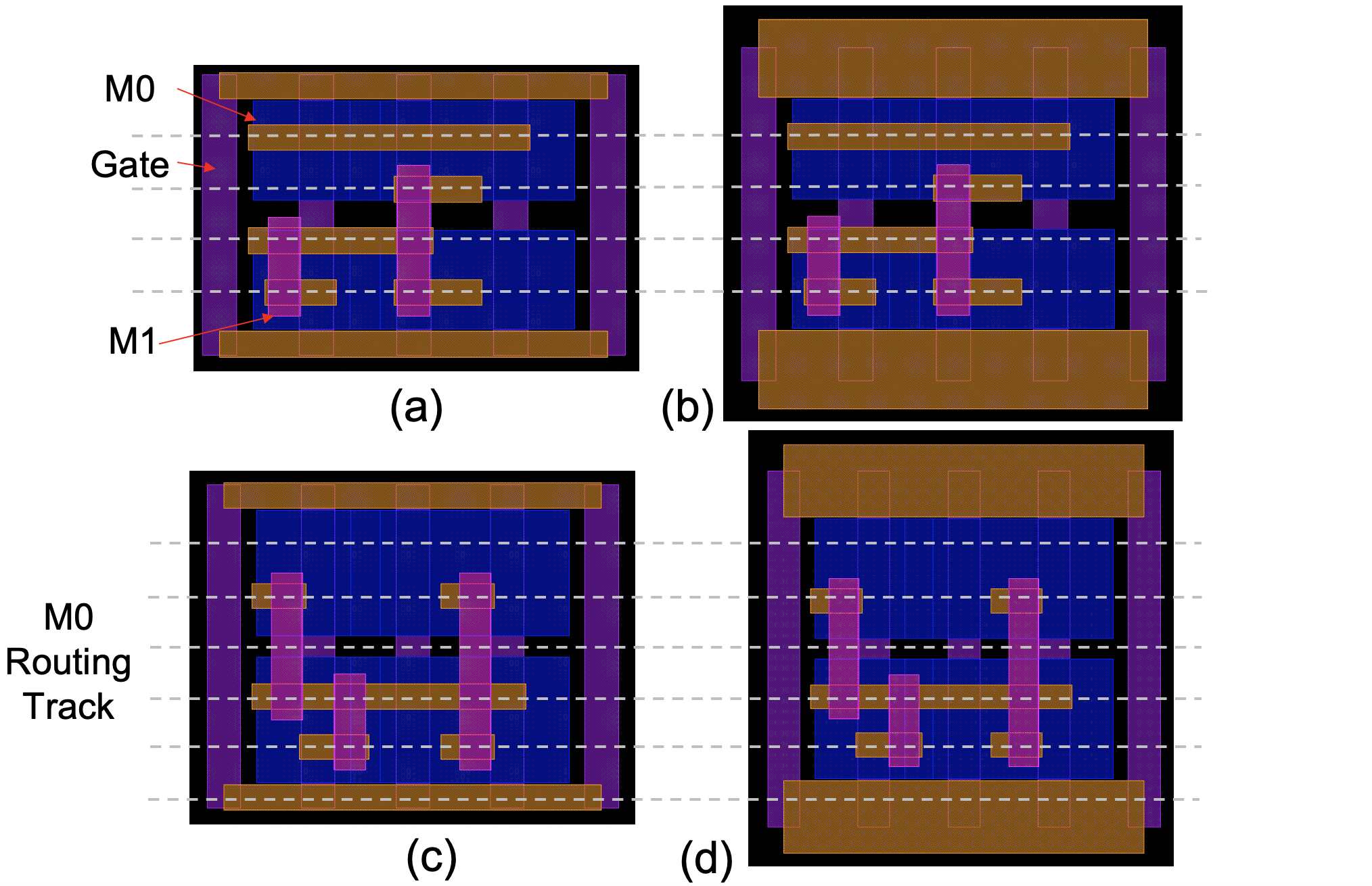}
\vspace{-0.2cm}
\caption{Example standard cells (AND2\_X1) in this work. 
The cells are generated by our SMT-based standard-cell layout
generation with the following parameters ($Fin$, $RT$, $PGpin$, $CH$): 
(a) {\em Lib1} ($2Fin$, $4RT$, $BPR$, $5T$); (b) {\em Lib2} ($2Fin$, $4RT$, $M0$, $6T$); 
(c) {\em Lib3} ($3Fin$, $5RT$, $BPR$, $6T$); and (d) {\em Lib4} ($3Fin$, $5RT$, $M0$, $7T$).} 
\vspace{-0.4cm}
\label{fig:stdcell}
\end{figure}

Our standard-cell layouts are generated using SPICE netlists, 
technology and design parameters from ~\cite{ChengKKK22}.
However, in PROBE3.0 we change two key parameters: metal pitch (MP) 
and power delivery network (PDN).
Instead of using MP, we define parameters for pitch values of each layer. 
Since M0, M1 and M2 layers are used for standard-cell layouts, 
we define M0P, M1P and M2P as pitches of M0, M1 and M2 layers, respectively.
Table~\ref{fig:stdcell} shows four layouts of AND2\_X1 cells with four 
parameter settings. The four standard-cell libraries ({\em Lib1}, {\em Lib2},
{\em Lib3} and {\em Lib4}) along with their corresponding parameter sets are 
used for our experiments in Section~\ref{sec:expt}. For our PPAC 
exploration, we generate 41 standard cells for each standard-cell library 
as shown in Table~\ref{tab:stdcelllist}.

\noindent
{\bf GDS/LEF Generation and LVS/PEX Flow.}
While \cite{ChengKKK22} only supports LEF generation for P\&R, 
PROBE3.0 generates standard-cell layouts in both GDS and LEF formats.
The GDS files are used to extract parasitics from standard-cell 
layouts and check LVS between layouts and schematics.
We use {\em Calibre}~\cite{Calibre} to check 
LVS and generate extracted netlists for standard cells with intra-cell 
RC parasitics.
Scripts for GDS/LEF generation and LVS/PEX flows are open-sourced in~\cite{PROBE3.0}.

\noindent
{\bf Library Characterization Flow.}
We perform library characterization to generate standard-cell 
libraries in the Liberty format.
The inputs to the flow are model cards 
for FinFET devices, Liberty template including PVT conditions,
and interconnect technology files. 
We use model cards from~\cite{FreePDK3}.
For the Liberty template, we define the PVT conditions, and the
capacitance and transition time indices of (7$\times$7) tables for 
electrical models (delay, output transition time, and power).
We use 5, 10, 20, 40, 80, 160, and 320$ps$ as the transition time indices.
For the input capacitance, we obtain the input pin 
capacitance $C_{inv}$ of an X1 inverter, then
multiply this value by  
predefined multipliers, 2, 4, 8, 16, 24, 32, and 64. 
For characterization, we use the PVT corner ($TT, 0.7V, 25\degree C$).

\begin{table}[t]
\vspace{-0.2cm}
\caption{List of 41 standard cells per generated library.} 
\vspace{-0.5cm}
\scriptsize
\label{tab:stdcelllist}
\begin{center}
\begin{tabular}{|c|c|}
\hline
{\bf Cell List}             & {\bf Size}       \\\hline
Inverter (INV), Buffer (BUF) & X1, X2, X4, X8   \\\hline
2-input AND/OR/NAND/NOR (AND2/OR2/NAND2/NOR2) & X1, X2   \\\hline
3-input AND/OR/NAND/NOR (AND3/OR3/NAND3/NOR3) & X1, X2   \\\hline
4-input NAND/NOR (NAND4/NOR4) & X1, X2   \\\hline
2-1 AND-OR-Inverter (AOI21), 2-2 AND-OR-Inverter (AOI22) & X1, X2   \\\hline
2-1 OR-AND-Inverter (OAI21), 2-2 OR-AND-Inverter (OAI22) & X1, X2   \\\hline
D flip-flop (DFFHQN), D flip-flop with reset (DFFRNQ) & X1  \\\hline
2-input MUX/XOR (MUX2/XOR2), Latch (LHQ) & X1  \\\hline
\end{tabular}
\end{center}
\vspace{-0.6cm}
\end{table}

\vspace{-0.3cm}
\section{Power Delivery Network}
\label{sec:powerdelivery}
\vspace{-0.1cm}

We study PDN scaling boosters to showcase the DTCO and pathfinding capability 
of PROBE3.0. There are two key
challenges of traditional PDNs at advanced technologies: 
\begin{itemize}
    \item {\em High resistance of BEOL}~\cite{Lu17}: Elevated resistance in 
    BEOL layers exacerbates 
    IR drop issues, necessitating denser PDN topologies.
    \item {\em Routing overheads (routability)}~\cite{SuHSN02}: PDN occupies 
    routing resources that are shared with signal and clock distribution. The routability 
    and area density impact of PDN becomes more severe with denser PDN at advanced nodes.
\end{itemize}

\begin{figure}[t]
\center
\includegraphics[width=0.95\columnwidth]{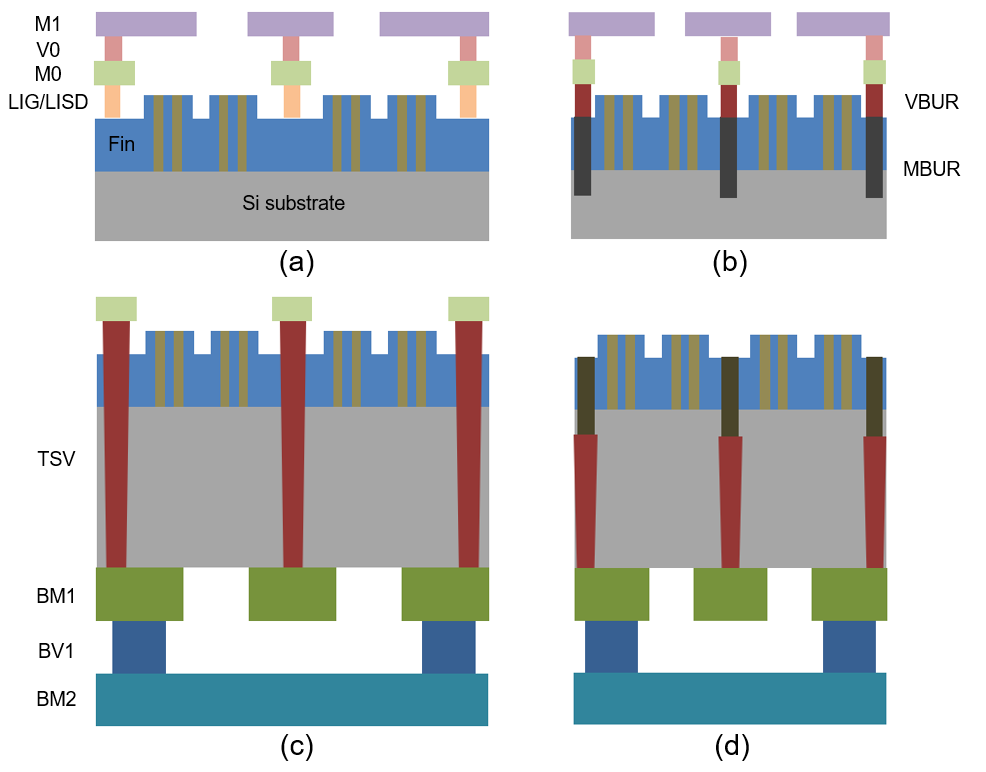}
\vspace{-0.2cm}
\caption{Cross-section view of four PDN options 
in the PROBE3.0 framework: (a) Frontside PDN ($P_{FS}$); (b) Frontside PDN with BPR ($P_{FB}$); 
(c) Backside PDN ($P_{BS}$); and (d) Backside PDN with BPR ($P_{BB}$).} 
\vspace{-0.3cm}
\label{fig:pdn}
\end{figure}

To overcome these challenges,
multiple foundries have begun implementing backside power delivery networks
(BSPDN) and buried power rails (BPR) as scaling boosters in their 
sub-5nm technologies. We use these scaling boosters, BSPDN and BPR, to demonstrate use of PROBE3.0.
We establish four options for $PDN$ parameter in the PROBE3.0 framework: 
(i) Frontside PDN without BPR ($P_{FS}$); 
(ii) Frontside PDN with BPR ($P_{FB}$); 
(iii) Backside PDN without BPR ($P_{BS}$);
and (iv) Backside PDN with BPR ($P_{BB}$).
Figure~\ref{fig:pdn} illustrates the four PDN configurations in the PROBE3.0 framework.

\vspace{-0.3cm}
\subsection{Frontside and Backside Power Delivery Network}
\label{subsec:fsbspdn}
\vspace{-0.1cm}

We have defined realistic structures 
for both frontside power delivery networks (FSPDN) and 
backside power delivery networks (BSPDN), and enabled
IR drop analysis within our framework.
Table~\ref{tab:pdn} shows the configurations for FSPDN and BSPDN.
Since BEOL layers with smaller pitches 
(e.g., 24nm-pitch layer) have high resistance, 
we add power stripes for every layer.
While the work 
of~\cite{ChengKKK22} has multiple options for FSPDN, the PROBE3.0 
framework has only one PDN structure for FSPDN. Instead, we add other 
options such as $P_{FB}$, $P_{BS}$ and $P_{BB}$. 
Furthermore, while the {\em Backside} option in~\cite{ChengKKK22} 
assumes no PDN at the frontside for the BSPDN option, we add power 
stripes at the backside for BSPDN in the PROBE3.0 framework to enable
IR drop analysis for BSPDN.

\begin{table}[t]
\vspace{-0.2cm}
\caption{PDN configurations for FSPDN and BSPDN. 
A pair of power (VDD) and ground (VSS) stripes are placed every pitch, 
while maintaining the spacing between VDD and VSS. 
{\em Density} denotes the percentage of routing tracks occupied by PDN.}
\vspace{-0.2cm}
\scriptsize
\label{tab:pdn}
\begin{center}
\begin{tabular}{|c|c|c|c|c|c|}
\hline
\multirow{2}{*}{{\bf PDN}} & \multirow{2}{*}{{\bf Layer}} & {\bf Pitch} & {\bf Width} & {\bf Spacing} & {\bf Density}\\
& & {\bf (um)} & {\bf (um)} & {\bf (um)} & {\bf (\%)}\\\hline
\multirow{4}{*}{FSPDN} & M3 & 1.08 & 0.012 & 0.508 & 4\\\cline{2-6}
& M4 & 1.152 & 0.032 & 0.544 & 11\\\cline{2-6}
& M5-M11 & 5.0 & 1.0 & 1.5 & 20\\\cline{2-6}
& M12-M13 & 4.32 & 1.8 & 0.36 & 100 \\\hline
BSPDN & BM1-BM2 & 4.32 & 1.8 & 0.36 & 100 \\\hline
\end{tabular}
\end{center}
\vspace{-0.5cm}
\end{table}

Figures~\ref{fig:pdn}(a) and (c) respectively show cross-section views of 
$P_{FS}$ and $P_{BS}$ options.
The $P_{FS}$ option has M0 power and ground pins for standard cells, which
connect to power stripes at the frontside of the die.
The $P_{BS}$ option uses the same M0 power and ground 
pins for standard cells but connects to power stripes at the backside of the die. 
For the $P_{BS}$ option, we employ two backside metal layers 
(BM1 and BM2) and one via layer (BV1) between the backside metal layers. 
The layer characteristics (width, pitch and spacing) are identical
to the top two layers (M12 and M13) of FSPDN. 
Additionally, the M0 pins of standard cells and BSPDN are connected 
using Through-Silicon Vias (TSVs). 
We assume nano-TSVs with 90nm~\cite{Sisto21} width for the $P_{BS}$ option,
and 1:10 width-to-height aspect ratio. For the $P_{BS}$ option, TSV insertions 
necessitate reserved spaces in front-end-of-line (FEOL) layers, including keepout 
margins surrounding the TSVs. To accommodate this, we insert 
{\em power tap cells} prior to standard-cell placement.

\vspace{-0.3cm}
\subsection{Frontside and Backside PDN with Buried Power Rail}
\label{subsec:bpr}
\vspace{-0.1cm}

In advanced nodes, power rails on BEOL metal layers can be 
``buried'' into FEOL levels with 
shallow-trench isolation (STI). Using deep trench and creating space between 
devices lowers the resistance of power rails.
In addition to the resistance benefits, standard-cell height (area) can be 
further reduced with  deep and narrow widths of power and ground pins.
Figures~\ref{fig:pdn}(b) and (d) respectively show cross-section views of 
FSPDN with BPR ($P_{FB}$) and BSPDN with BPR ($P_{BB}$) options. 
In the case of $P_{FB}$, connections between FSPDN and BPR are made through 
nano-TSVs with the same 90nm width as in the $P_{BS}$ option (but, with 1:7
aspect ratio). These nano-TSVs also necessitate insertion of reserved spaces. 

\vspace{-0.3cm}
\subsection{Power Tap Cell Insertion}
\label{subsec:powertap}
\vspace{-0.1cm}

Although use of BSPDN and BPR can reduce area and
mitigate IR drop problems, connecting frontside layers to BSPDN and/or BPR 
remains a critical challenge. 
To establish ``tap'' connections from frontside metals to 
BPR, or from backside to frontside metals, 
space must be reserved on device layers -- e.g., 
\cite{PrasadNDZ19} proposes power tap cells for the connection
between BPR to MINT (M0) layers.
More frequent ``taps'' will mitigate IR drop problems, but occupy
more placement area.
In PROBE3.0, we define two types of power tap cells for 
the $P_{FB}$ and $P_{BS}$ options.
Tap cells for $P_{FB}$ connect BPR to M1, and tap cells for $P_{BS}$ connect BM1 to M0.
By contrast, $P_{FS}$ and $P_{BB}$ do not require power tap cells.

\begin{figure}[t]
\center
\includegraphics[width=0.95\columnwidth]{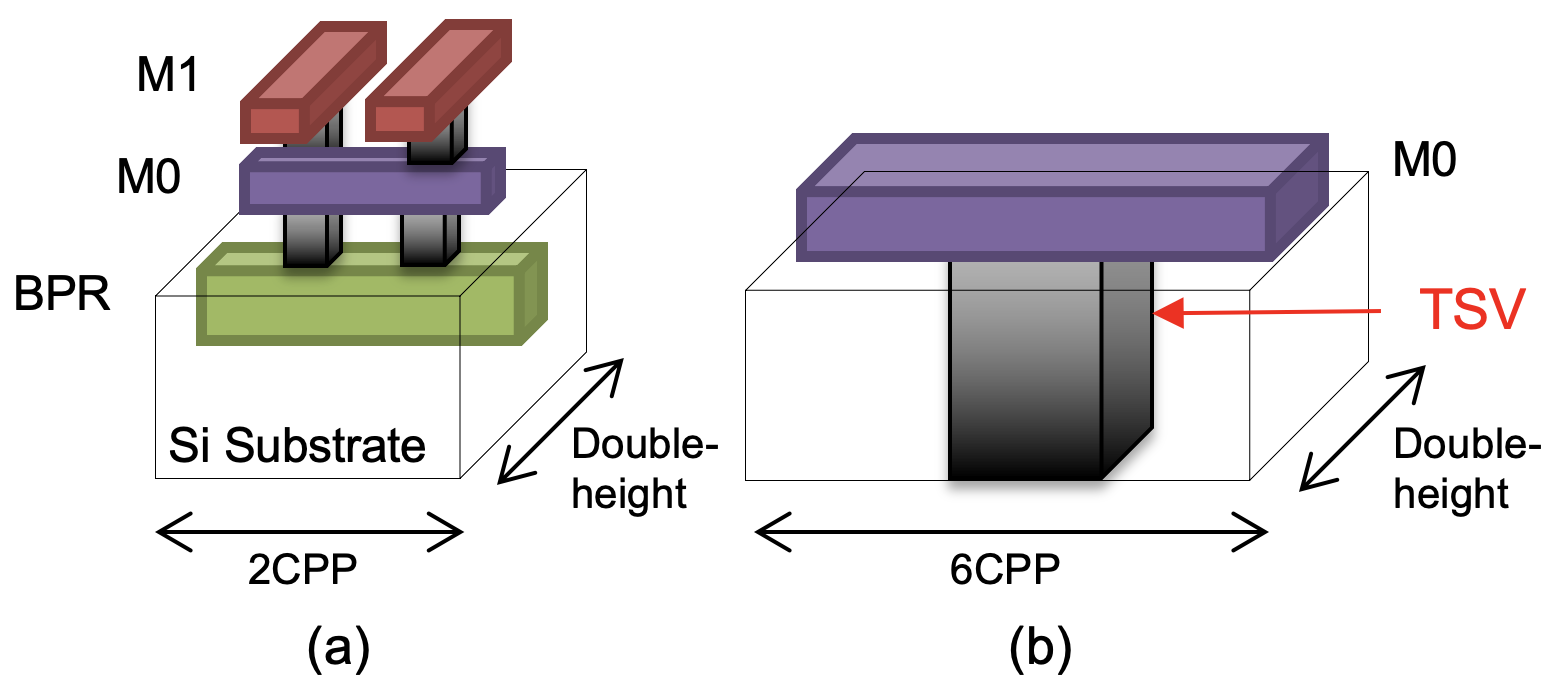}
\vspace{-0.3cm}
\caption{Power tap cells for (a) $P_{FB}$ and (b) $P_{BS}$.} 
\vspace{-0.5cm}
\label{fig:powertapcell}
\end{figure}

\noindent
{\bf Power Tap Cell Structure.}
Figure~\ref{fig:powertapcell}(a) shows a structure of power tap cells for $P_{FB}$.
Double-height power tap cells for $P_{FB}$ have 2CPP cell width.
The connection between BPR and M0 is through a 1$\times$2 via array, and the two M1 metals
are aligned with M1 vertical routing tracks.
There are also two types of power tap cells for $P_{FB}$ 
according to starting power and ground pins: 
power/ground pins on the double-height power tap cells are ordered as
Power-Ground-Power (VDD-VSS-VDD) or Ground-Power-Ground (VSS-VDD-VSS).
On the other hand, Figure~\ref{fig:powertapcell}(b) shows a structure of
power tap cells for $P_{BS}$.
While power tap cells for $P_{FB}$ have 2CPP width, double-height power tap cells
for $P_{BS}$ have 6CPP width due to the $\sim$90nm width of 
nano-TSVs~\cite{Sisto21}.
We also assume a 50nm keepout spacing around nano-TSVs. 
Similar to power tap cells for $P_{FB}$, there are two types of double-height 
power tap cells for $P_{FB}$,
Power-Ground-Power and Ground-Power-Ground.

\noindent
{\bf Power Tap Cell Insertion Scheme.}
Power tap cell insertion affects routability and IR drop, and hence 
affect PPAC of designs.
In this work, we define five tap cell insertion pitches and two power tap
insertion schemes, as follows.

\begin{itemize}
    \item $I_{pitch}$: 24, 32, 48, 96 and 128CPP
    \item $I_{scheme}$: {\em Column} and {\em Staggered}
\end{itemize}

\begin{figure}[t]
\center
\includegraphics[width=1.0\columnwidth]{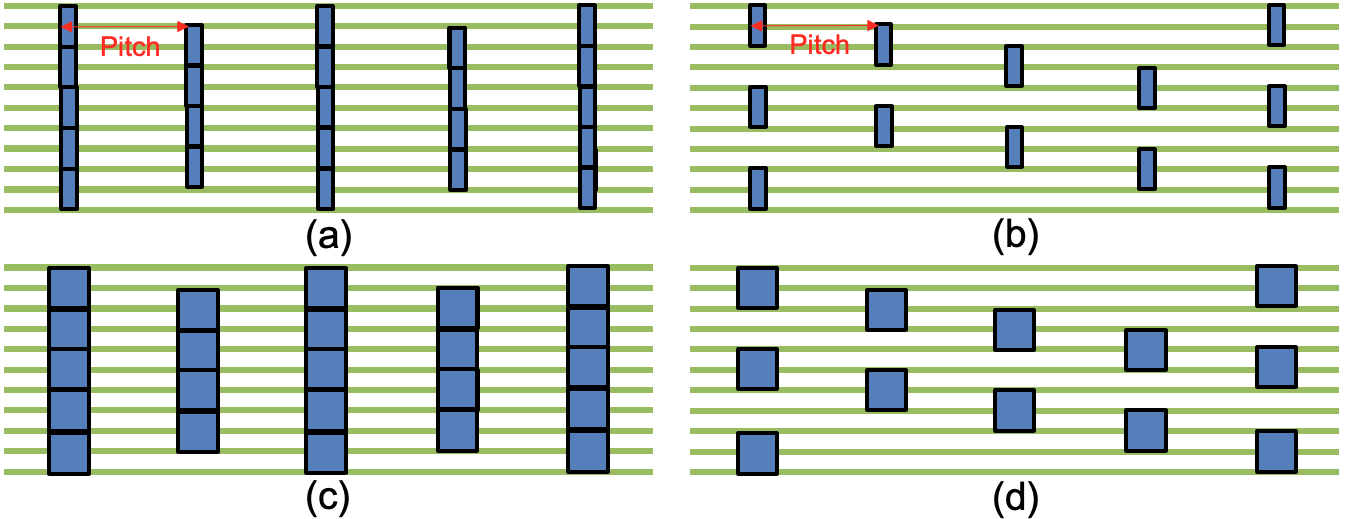}
\vspace{-0.4cm}
\caption{Four power tap cell insertion results. (a) Power tap cells for $P_{FB}$
(2CPP width) with {\em Column}; (b) power tap cells for $P_{FB}$ with 
{\em Staggered}; (c) power tap cells for $P_{BS}$ (6CPP width) with {\em Column}; and 
(d) power tap cells for $P_{BS}$ with {\em Staggered}.} 
\vspace{-0.5cm}
\label{fig:powertap}
\end{figure}

$I_{pitch}$ and $I_{scheme}$ denote tap cell insertion pitch and tap cell 
insertion scheme, respectively.
Tap cell insertion scheme {\em Column} places double-height power tap cells on
every two placement rows with the given tap cell pitch.
Conversely, tap cell insertion scheme {\em Staggered} places 
double-height power tap cells on
every four placement rows with the given tap cell pitch.
Figure~\ref{fig:powertap} shows four power tap cell insertion results for 
$P_{FB}$ and $P_{BS}$ with {\em Column} and {\em Staggered} insertion schemes.

\vspace{-0.3cm}
\subsection{IR Drop Analysis Flow}
\label{subsec:irflow}
\vspace{-0.1cm}

\begin{figure}[t!]
\center
\includegraphics[width=1.0\columnwidth]{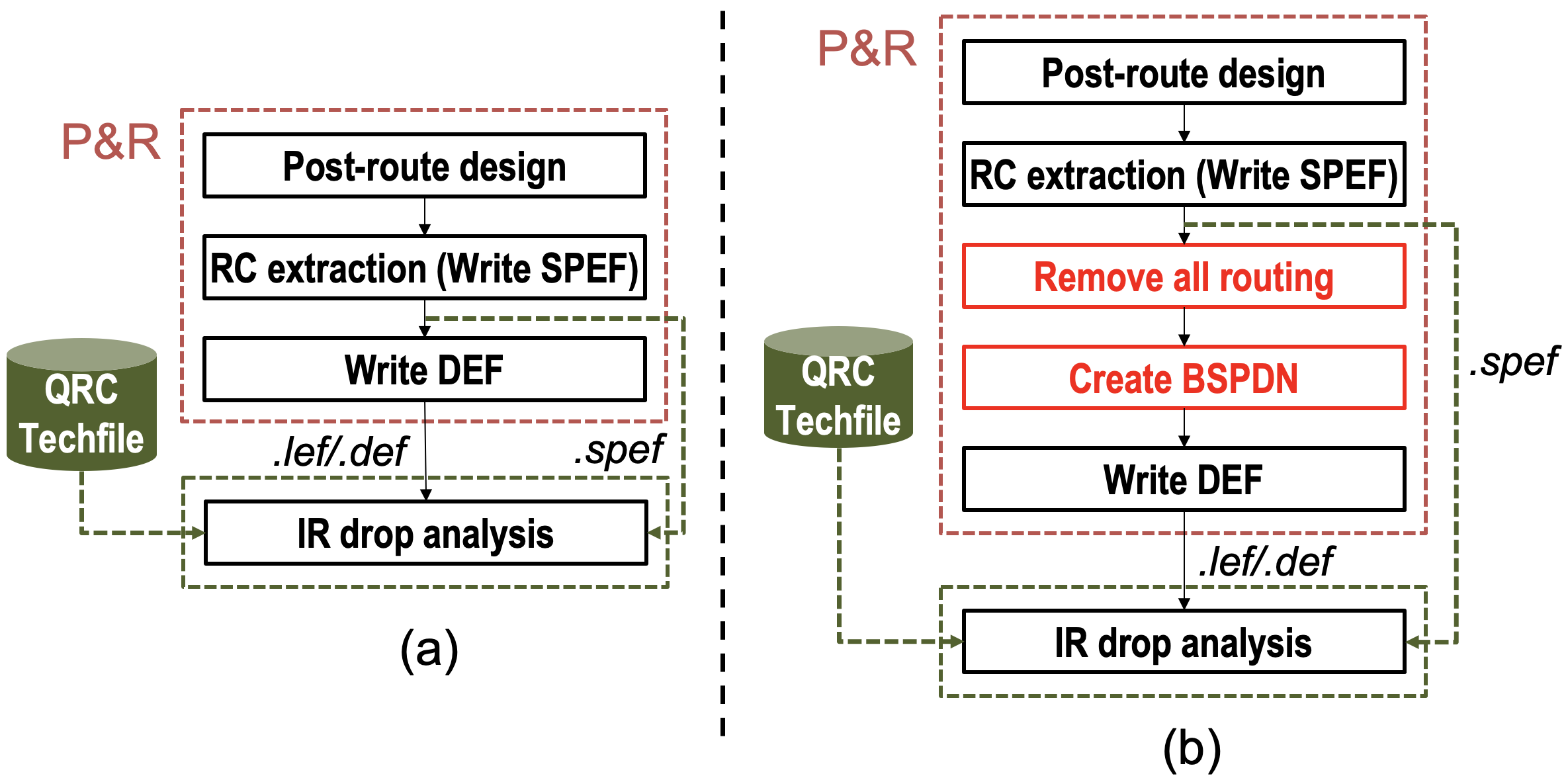}
\vspace{-0.3cm}
\caption{IR drop analysis flow for (a) FSPDN and (b) BSPDN. For the IR drop flow for BSPDN, 
we delete all the signal and clock routing after P\&R and build power stripes for BSPDN.} 
\vspace{-0.5cm}
\label{fig:ir_bspdn}
\end{figure}

We develop two 
IR drop analysis flows for FSPDN and BSPDN.
Figure~\ref{fig:ir_bspdn}(a) presents our IR drop analysis flow for FSPDN. 
After P\&R, we generate DEF and SPEF files for routed designs using 
a commercial P\&R tool to perform standalone vectorless dynamic 
IR drop analysis. Additionally, an interconnect technology file (QRC techfile)
is needed for RC extraction as input for the IR drop analysis flow.
In contrast, Figure~\ref{fig:ir_bspdn}(b) depicts our IR drop analysis flow 
for BSPDN. After P\&R, we only create a SPEF file from routed designs.
We then remove all routed signals and clocks from the P\&R database 
and construct new power stripes for BSPDN. Since the standalone IR drop 
analysis tool obtains power stripe information from a DEF file, we generate
a DEF file after creating power stripes on the backside. 
There are two backside metal layers, BM1 and BM2. 
When creating PDN on backside metal layers, we consider M1 as BM1 and M2 as BM2,
respectively.
For RC extraction with BSPDN, the QRC techfile must be scaled for 
backside metals since we assume BM1 and BM2 have the same pitches as M12 and M13.
Full details are visible in open-source scripts at \cite{PROBE3.0}.
%XXX Please add the citation.

\vspace{-0.3cm}
\section{Enhanced Artificial Designs for PPAC Exploration}
\label{sec:ang}
\vspace{-0.1cm}

The use of specific real designs in DTCO and PPAC exploration
can bring risk of biases and incorrect decisions regarding technology
configurations (e.g., cell architecture or BEOL stack).
To avoid such biases, the 
PROBE1.0~\cite{KahngKLL18} bases its routability assessment
on a mesh-like netlist topology, and PROBE2.0~\cite{ChengKKK22} similarly
uses a knight's tour-based topology. 
However, these artificial topologies have two main limitations as we bring ``PP''
aspects of PPAC into the picture.
First, they are highly regular and cannot capture a wide range of circuit types.
Second, they do not mimic timing and power properties of
real netlists, as they target routability assessment without regard
to timing path structure. 

PROBE3.0 overcomes these limitations
by generating artificial but realistic netlists with 
the {\em Artificial Netlist Generator} (ANG) of ~\cite{KimLMK22}\cite{ANG},
for use in PPAC studies. 
We use the six topological parameters of ANG (see
Table~\ref{tab:angparamdef})
to generate and explore circuits with various sizes, interconnect complexity, 
routed wirelengths and timing.
Moreover, we apply machine learning (AutoML) to improve the match of generated
artificial netlists to targeted (real) netlists. 

\begin{table}[t!]
\vspace{-0.2cm}
\caption{Definition of topological parameters in ANG   
~\cite{KimLMK22}\cite{ANG}.}
\vspace{-0.4cm}
\scriptsize
\label{tab:angparamdef}
\begin{center}
\begin{tabular}{|m{1.4cm}|m{6.7cm}|}
\hline
{\bf Parameter} & {\bf Definition}\\\hline
$N_{inst}$ ($T_1$) & Number of instances.\\\hline
$N_{prim}$ ($T_2$) & Number of primary inputs/outputs.\\\hline
$D_{avg}$ ($T_3$) & Average net degree. The net degree of a net 
    is the number of terminals of the net.\\\hline
$B_{avg}$ ($T_4$) & Average size of net bounding box. The placed 
    (or routed) layout is divided into a bin grid where each bin contains 
    $\sqrt{N_{inst}}$ instances.\\\hline
$T_{avg}$ ($T_5$) & Average depth of timing paths. 
    The depth of a given timing endpoint is the maximum number 
    of stages in any fanin combinational path of that endpoint.
    $T_{avg}$ is the average of all endpoint depths.\\\hline
$S_{ratio}$ ($T_6$) & Ratio of the number of sequential cells to 
    the total number of cells. $S_{ratio}$ equals to number of sequential 
    cells over total number of instances.\\\hline
\end{tabular}
\end{center}
\vspace{-0.5cm}
\end{table}

\vspace{-0.3cm}
\subsection{Comparison of ANG and Real Designs}
\label{subsec:angcomp}
\vspace{-0.1cm}

In this subsection, we study
four real designs from OpenCores~\cite{OpenCores}
and the corresponding artificial netlists generated 
by ANG~\cite{KimLMK22}. 
Each design is taken through commercial logic synthesis 
and P\&R tools~\cite{DC}\cite{ICC2} in the PROBE3.0 technology,
to obtain a final-routed layout. 
For AES, JPEG, LDPC and VGA,
we respectively use target clock periods of 
0.2ns, 0.2ns, 0.6ns and 0.2ns, and utilizations of  
0.7, 0.7, 0.2 and 0.7. 
We then extract the six topological parameters from the routed designs 
and use these parameters to generate artificial netlists with ANG. 

We introduce a {\em Score} metric to quantify similarity between artificial 
and real netlists, as defined in Equation~(\ref{eq:score}).

\vspace{-0.2cm}
\begin{equation}
\begin{aligned}
Score = \Pi_{i=1}^{N}{\max(\frac{T^{target}_i}{T^{out}_i}, \frac{T^{out}_i}{T^{target}_i})}
\label{eq:score}
\end{aligned}
\end{equation}
\begin{conditions}
T^{target}_i & $T_i$ in target parameter set \\
T^{out}_i & $T_i$ of output parameter set \\   
N & number of parameters ($N=6$)
\end{conditions}
\vspace{-0.2cm}

In Equation~(\ref{eq:score}), target and output parameters 
are elements $T^{target}_i$ and $T^{out}_i$ of the target and output parameter sets.
For each parameter, we calculate the discrepancy (ratio) between target and output values.
The $Score$ value is the  product of these ratios. 
Ideally, if output parameters are exactly the same as target parameters, 
$Score$ is 1.
Larger values of $Score$ indicate greater discrepancy between ANG-generated netlists 
and the target netlists.

\begin{table}[t]
\vspace{-0.3cm}
\caption{Topological parameters for real netlists from
OpenCores~\cite{OpenCores} and artificial netlists generated 
by~\cite{KimLMK22}. Design names followed by $*$ indicate
ANG-generated artificial netlists.}
\vspace{-0.3cm}
\scriptsize
\label{tab:angparam}
\begin{center}
\begin{tabular}{|c|c|c|c|c|c|c|c|}
\hline
\multirow{2}{*}{{\bf Design}} &  \multicolumn{6}{c|}{{\bf Parameters}} & \multirow{2}{*}{{\bf Score}}\\\cline{2-7}
 & $N_{inst}$ & $N_{prim}$ & $D_{avg}$ & $B_{avg}$ & $T_{avg}$& $S_{ratio}$ & \\\hline\hline
AES   & 12318 & 394  & 3.28 & 0.55 & 7.98  & 0.04 & - \\\hline
JPEG  & 70031 & 47   & 3.09 & 0.21 & 10.36 & 0.07 & - \\\hline
LDPC  & 77379 & 4102 & 2.85 & 1.00 & 12.94 & 0.03 & - \\\hline
VGA   & 60921 & 185  & 3.71 & 0.42 & 8.25  & 0.28 & - \\\hline\hline
AES*  & 10371 & 394  & 3.28 & 0.79 & 5.19  & 0.13 & 8.53 \\\hline
JPEG* & 63185 & 47   & 3.16 & 0.70 & 6.97  & 0.15 & 12.03 \\\hline
LDPC* & 58699 & 4106 & 3.10 & 0.78 & 6.96  & 0.13 & 14.8 \\\hline
VGA*  & 64412 & 188  & 3.32 & 0.26 & 6.39  & 0.25 & 2.8 \\\hline
\end{tabular}
\end{center}
\vspace{-0.5cm}
\end{table}

Table~\ref{tab:angparam} shows the input parameters, extracted parameters 
and $Score$ metric in our comparison of real and artificial designs.
The causes of discrepancy are complex, e.g., 
\cite{KimLMK22} has steps that heuristically adjust average depths of 
timing paths $T_{avg}$ and the ratio of sequential cells $S_{ratio}$.
Also, performing P\&R will change the number of instances $N_{inst}$, 
the average net degree $D_{avg}$,
and the routing which determines $B_{avg}$. 
Hence, it is difficult to identify the input parameterization of ANG
that will yield artificial netlists whose post-route properties match those of
(target) real netlists. We use machine learning to address this challenge. 

\vspace{-0.3cm}
\subsection{Machine Learning-Based ANG Parameter Tuning}
\label{subsec:angautoml}
\vspace{-0.1cm}

\begin{figure}[t]
\center
\includegraphics[width=1.0\columnwidth]{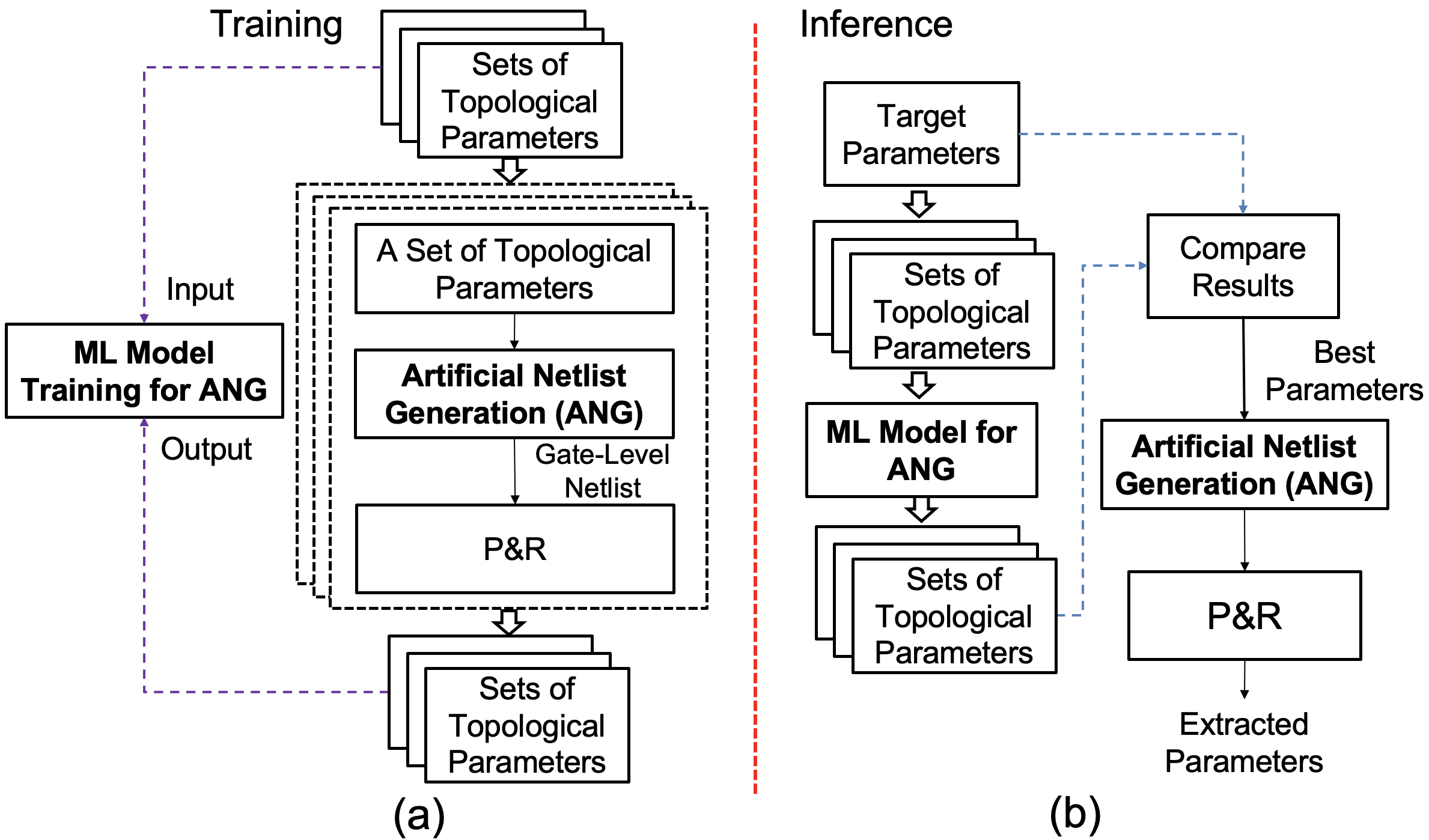}
\vspace{-0.3cm}
\caption{ML-based parameter tuning for ANG.} 
\vspace{-0.3cm}
\label{fig:mlang}
\end{figure}

We improve the realism of generated artificial netlists
with ML-based parameter tuning for ANG.
Figure~\ref{fig:mlang}(a) shows the training flow in the
parameter tuning. 
First, to generate training data, we sweep the six ANG input parameters to 
generate 21,600 combinations of input parameters, as described in 
Table~\ref{tab:traintest}. Second, we use ANG with 
these input parameter combinations to generate artificial gate-level netlists.
Third, we perform P\&R with the (21,600) artificial netlists and extract the output parameters.
The extracted output parameters are used as output labels 
for the ML model training.
We use the open-source H2O AutoML package~\cite{H2OAutoML} (version 3.30.0.6) to predict 
the output parameters; the {\em StackedEnsemble\_AllModels} model consistently
returns the best model.
The model training is a one-time overhead which took 4 hours using an Intel Xeon 
Gold 6148 2.40GHz server. Executing commercial P\&R required just over 7 days in our academic
lab setting, and is again a one-time overhead.\footnote{The average P\&R runtime on our 21,600
ANG netlists is 0.4 hours on an Intel Xeon Gold 6148 2.40GHz server. The data generation used 
50 concurrently-running licenses of the P\&R tool, with each job running single-threaded. 
(21,600 $\times$ 0.4 / 50 / 24 $\sim =$ 7.2 days. With multi-threaded runs, we estimate 
that data generation would have taken 3 to 4 days.)}

Figure~\ref{fig:mlang}(b) shows our inference flow. 
First, we define ranges around the target parameter and sweep the parameters to generate 
multiple combinations of input parameters as candidates, which are
shown in Table~\ref{tab:traintest}.
Second, we use our trained model to predict the output parameters from each input
parameter combination.
Note that although there are 12.3M combinations as specified in the rightmost 
two columns of Table~\ref{tab:traintest}, this step requires less than 10 minutes on
an Intel Xeon Gold 6148 2.40GHz server.\footnote{11 $\times$ 11 $\times$ 21 $\times$ 21 $\times$ 11 $\times$ 21 = 12,326,391. We apply simple filtering based on lower and
upper bounds, to avoid parameter values for which ANG does not work properly. Specifically, 
parameter values are restricted to be within:
$0 < B_{avg} \leq 1.0$; 
$0 < S_{ratio} \leq 1.0$; 
$1 < D_{avg} <  2.6$; and
$3 < T_{avg}$.  For example, the AES testcase then has $\sim$3M input parameter
combinations, and predicting output parameters for all of these takes 441 seconds of runtime.}
Third, we calculate a predicted $Score$ per each input parameter combination,
and then choose the parameter combination with lowest predicted $Score$.
Finally, we use ANG and the chosen parameter combination to generate an artificial 
netlist for P\&R and PPAC explorations.

Table~\ref{tab:mlang} shows the benefit from ML-based ANG parameter tuning.
Columns 2-5 show parameters from real netlists, which we use as target 
parameters.
The trained ML model and the inference flow produce the tuned 
parameters for ANG shown in Columns 6-9 of the table, and
corresponding results are shown in Columns 10-13. 
The average {\em Score} decreases from to 4.89 from the
original value of 8.87 for ANG without ML-based parameter tuning 
(Table~\ref{tab:angparam}).

\begin{table}[tb]
\vspace{-0.1cm}
\caption{Parameter sets for training and testing.
We train our ML model with ANG input parameters and post-P\&R output 
parameters. The total number of datapoints is 
$4\times6\times6\times5\times5\times6 = 21600$. Testing is performed in the 
ranges around given target parameters, according to the step sizes.}
\vspace{-0.1cm}
\scriptsize
\label{tab:traintest}
\begin{center}
\begin{tabular}{|c||c|c|c|}
\hline
\multirow{2}{*}{{\bf Parameter}} & \multirow{2}{*}{{\bf Training Value}} & \multicolumn{2}{c|}{{\bf Testing Value}}\\\cline{3-4}
 & & {\bf Range} & {\bf Step}\\\hline
$N_{inst}$ ($T^{in}_1$) & 10000, 20000, 40000, 80000 & $T^{target}_1 \pm 500$ & 100\\\hline
$N_{prim}$ ($T^{in}_2$) & 100, 200, 500, 1000, 2000, 4000 & $T^{target}_2 \pm 5$ & 1\\\hline
$D_{avg}$ ($T^{in}_3$) & 1.8, 2.0, 2.2, 2.4, 2.6 & $T^{target}_3 \pm 0.2$ & 0.02\\\hline
$B_{avg}$ ($T^{in}_4$) & 0.70, 0.75, 0.80, 0.85, 0.90, 0.95 & $T^{target}_4 \pm 0.2$ & 0.02 \\\hline
$T_{avg}$ ($T^{in}_5$) & 6, 8, 10, 12, 14, 16 & $T^{target}_5 \pm 10$ & 2\\\hline
$S_{ratio}$ ($T^{in}_6$) & 0.2, 0.4, 0.6, 0.8, 1.0 & $T^{target}_6 \pm 0.2$ & 0.02\\\hline
\end{tabular}
\end{center}
\vspace{-0.1cm}
\end{table}

\begin{table*}[t]
\vspace{-0.3cm}
\caption{Topological parameters for target, input and output netlists. The design names followed by $**$ indicate ANG-generated
artificial netlists with ML-based ANG parameter tuning.}
\vspace{-0.3cm}
\scriptsize
\label{tab:mlang}
\begin{center}
\begin{tabular}{|c||c|c|c|c||c|c|c|c||c|c|c|c|}
\hline
\multirow{2}{*}{{\bf Parameter}}& \multicolumn{4}{c||}{{\bf Parameters of Target Netlists}}& \multicolumn{4}{c||}{{\bf ANG Input Parameters (ML Inference)}}& \multicolumn{4}{c|}{{\bf Parameters from Artificial Netlists}}\\\cline{2-13}
 & {\bf AES} & {\bf JPEG} & {\bf LDPC} & {\bf VGA}& {\bf AES} & {\bf JPEG} & {\bf LDPC} & {\bf VGA}& {\bf AES**} & {\bf JPEG**} & {\bf LDPC**} & {\bf VGA**}\\\hline
$N_{inst}$ & 12318 & 70031 & 77379 & 60921 & 12718 & 69531 & 76979 & 60421 & 10200 & 64296 & 64796 & 65113 \\\hline
$N_{prim}$ & 394   & 47    & 4102  & 185   & 390   & 42    & 4106  & 199   & 394   & 46    & 4110 & 202   \\\hline
$D_{avg}$  & 3.28  & 3.09  & 2.85  & 3.71  & 3.40  & 3.10  & 3.03  & 3.53  & 3.26  & 3.13  & 3.18 & 3.30  \\\hline
$B_{avg}$  & 0.55  & 0.21  & 1.00  & 0.42  & 0.49  & 0.31  & 1.98  & 0.28  & 0.72  & 0.21  & 0.73 & 0.36  \\\hline
$T_{avg}$  & 7.98  & 10.36 & 12.94 & 8.25  & 13.98 & 18.36 & 20.94 & 12.25 & 8.01  & 9.29  & 11.64 & 8.54  \\\hline
$S_{ratio}$& 0.04  & 0.07  & 0.03  & 0.28  & 0.01  & 0.27  & 0.01  & 0.16  & 0.11  & 0.20  & 0.13 & 0.16  \\\hline
$Score$ & - & - & - & - & - & - & - & - & 4.39 & 3.59 & 2.77 & 8.81 \\\hline
\end{tabular}
\end{center}
\vspace{-0.5cm}
\end{table*}

\begin{figure}[t]
\center
\includegraphics[width=1.0\columnwidth]{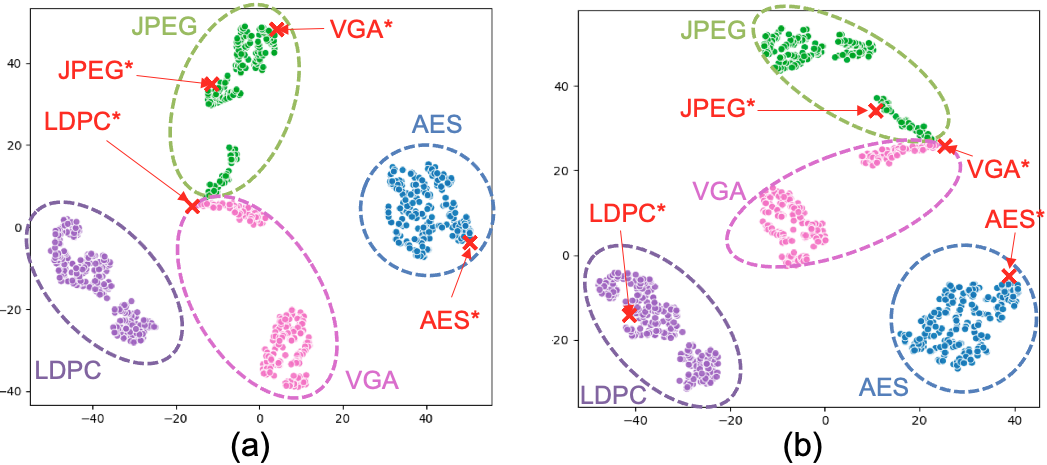}
\vspace{-0.5cm}
\caption{Comparison between real and artificial designs by t-SNE~\cite{MaatenH08}.
(a) t-SNE visualization for real and artificial (ANG) designs {\bf without} our 
parameter tuning flow. (b) Real and artificial (ANG) designs {\bf with}
our ML-based parameter tuning flow.  Design names followed by 
$*$ indicate artificial designs.
}
\vspace{-0.3cm}
\label{fig:tsne}
\end{figure}

The ML-enabled improvement of realism in ANG netlists can be seen
using t-SNE visualization~\cite{MaatenH08} from P\&R results.
We perform P\&R for the four real designs by sweeping initial utilization 
from 0.6 to 0.8 with a 0.01 step size, and target clock period from 
0.15 to 0.25ns with a 0.01ns step size; this results in 21 $\times$ 11 = 231
P\&R runs. (For LDPC, we sweep utilization from 0.1 to 0.3 with a 0.01 step size,
and clock period from 0.55 to 0.65ns with a 0.01ns step size.)
We then perform P\&R for artificial netlists with and without our parameter tuning 
flow,  with 0.7 utilization (0.2 for LDPC) and 0.2ns (0.6ns LDPC) target clock period. 
Figure~\ref{fig:tsne} shows t-SNE visualization\footnote{For t-SNE visualization,
we collect ten features from P\&R results: Number of instances, number of nets, 
number of primary input/output pins, average fanout, number of sequential 
cells, wirelength, area, number of design rule violations, worst negative slack, 
total negative slack, and number of failing endpoints.}
of the real and artificial designs. 
The 231 real datapoints per design form well-defined clusters. 
In Figure~\ref{fig:tsne}(a), the datapoints of the artificial AES and JPEG 
designs are located in the corresponding designs' clusters. However,
the artificial LDPC and VGA designs are not close to the corresponding clusters of
real designs. 
By contrast, Figure~\ref{fig:tsne}(b) shows that with our ML-based ANG parameter
tuning, datapoints of all four artificial designs are located 
within the corresponding clusters of real designs.
This suggests that the ML-based ANG parameter tuning helps create artificial 
netlists that better match targeted design parameters -- including parameters
that are relevant to PPAC exploration.

\vspace{-0.3cm}
\section{Cell Width-Regularized Placements for More Realistic Routability Assessment}
\label{sec:enhancedcw}
\vspace{-0.1cm}

Recall that in the PROBE approach, routability (``AC'') is evaluated using 
the $K$-threshold ($K_{th}$) metric ~\cite{KahngKLL18}. 
That is, given a placed netlist, routing difficulty is gradually increased 
by iteratively swapping random pairs of neighboring 
instances. The cell-swaps progressively ``tangle'' the placement until it
becomes unroutable ($> 500$ DRCs post-detailed routing). 
The number of swaps $K$ -- expressed as a multiple of the instance count --
at which routing fails is the $K_{th}$ metric. Larger $K_{th}$ implies greater 
routing capacity or intrinsic routability.

Both PROBE1.0~\cite{KahngKLL18} and PROBE2.0~\cite{ChengKKK22} 
enable the study of real netlists through the concept of a {\em cell width-regularized} 
placement. In this approach, combinational cells are inflated (by LEF modification)
to match the maximum width among all the combinational cells in the cell library.
This process, called {\em cell width-regularization}, 
prevents illegal placements (i.e., cell overlaps due to varying widths) 
from arising due to neighbor-swaps during $K_{th}$ evaluation.
Unfortunately, while cell width-regularization permits real designs to be 
placed and then tangled by random neighbor-swaps, it also forces low utilizations 
that harm the realism of the study. (Moreover, high whitespace leads to high $K_{th}$
values that require more P\&R runs to determine.)

We now describe a {\em clustering-based cell width-regularization} methodology 
that generates placements with realistic utilizations, based on real designs.
Our experiments in Section~\ref{subsec:kth}
show that clustering-based cell width-regularization
obtains the same $K_{th}$ rank-ordering of design enablements, with less P\&R expense, 
than the previous cell width-regularization approach.

\begin{figure}[t]
\center
\includegraphics[width=1.0\columnwidth]{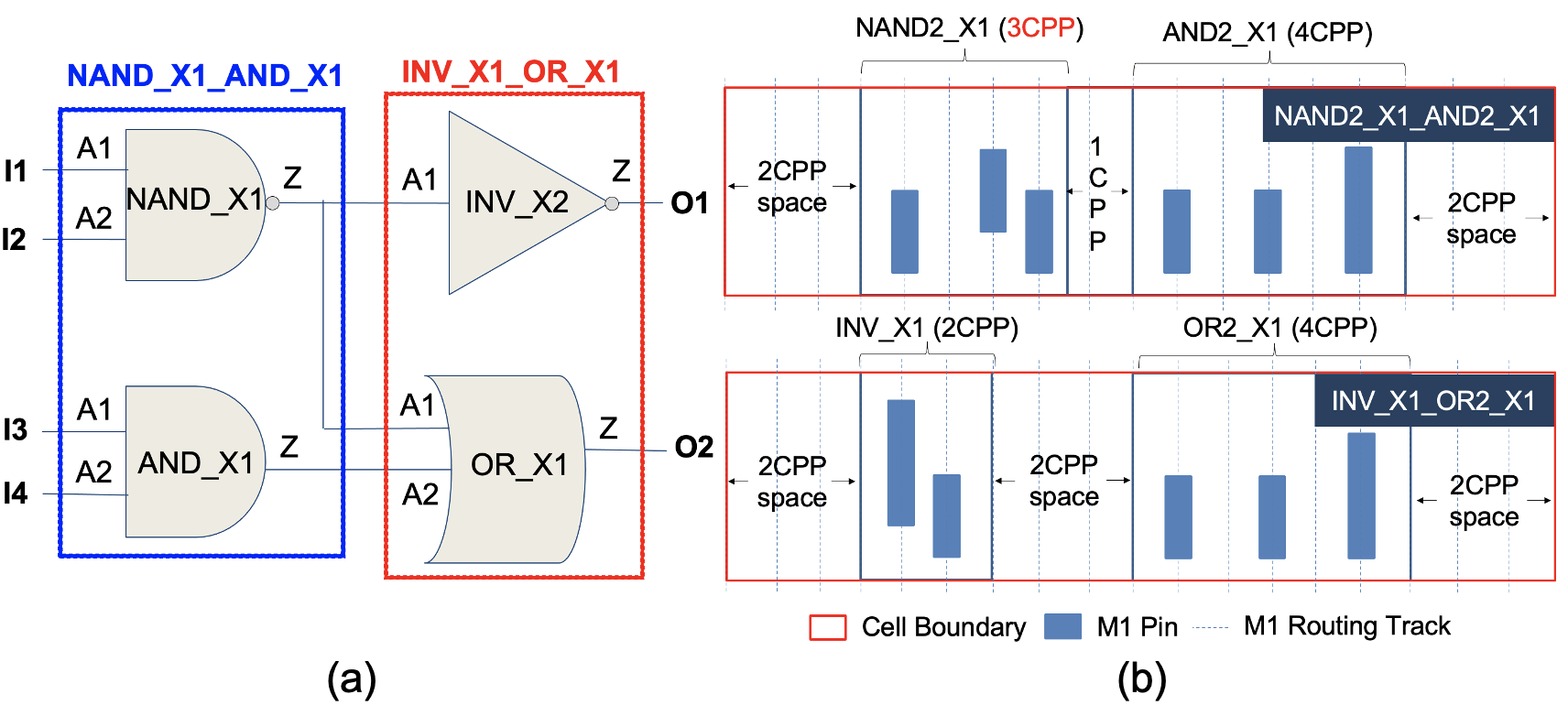}
\vspace{-0.5cm}
\caption{Two example clustered cells 
NAND\_X1\_AND\_X1, and INV\_X1\_OR\_X1 in
clustering-based cell width-regularization.
(a) Schematic view, and (b) physical layout view assuming {\em Lib2}.
Here, the maximum clustered cell width $w_{max}$ is 12CPP.}
\vspace{-0.2cm}
\label{fig:toy_example}
\end{figure}

\vspace{-0.3cm}
\subsection{Clustering-Based Cell Width-Regularization}
\label{subsec:Clustered}
\vspace{-0.1cm}

We propose {\em clustering-based cell width-regularization} 
using bottom-up hypergraph clustering, as detailed 
in Algorithm~\ref{alg:clustering}. 
In the following, we refer to standard cells of the original netlist as 
$cells_{orig}$, and (clustered) cells of the clustered netlist as 
$cells_{clustered}$.  

\begin{algorithm}[!t]
\scriptsize
    \caption{Cell width-regularization by clustering.}
    \begin{algorithmic}
    \State {\bf Inputs:} Hypergraph $H(V,E,W)$, Maximum cell width $w_{max}$, Number of iterations $N_{iter}$
    \State {\bf Outputs:}, Clustered hypergraph $H_{out}(V_{out}, E_{out}, W_{out})$
    \end{algorithmic}
    \begin{algorithmic}[1]
    \State $N_{cluster} \leftarrow |V|$ 
    \State Hypergraph at iteration $0$, $H_0 (V_0,E_0,W_0) \leftarrow H(V,E,W)$ 
    \For{$k \leftarrow 0$; $k < N_{iter}$ ; $k++$} 
    \State $V_{ordered} \leftarrow$ Sorted $V_{k}$ in increasing order of $W_k$
    \State $visited[v] \leftarrow false$ $\forall v \in V_k$
    \State Cluster assignments, $cmap[v] \leftarrow v ~\forall v \in V_k$
    \State Clustered cell widths, $W_{k+1} \leftarrow W_k$
    \For{$v_i \in V_{ordered}$}
        \If{$visited[v_i] == true$ or $v_i$ is a sequential cell}
        \State continue
        \EndIf
        \State $V_{neighbor} \leftarrow$ Find adjacent vertices of $v_i$
        \State Best cluster score, $\phi _{best} \leftarrow 0$; Best cluster candidate, $v_{best} \leftarrow -1$
        \For{$v_j$ in $V_{neighbor}$}
            \If{$W_k[{v_i}] + W_{k+1}[cmap[{v_j}]] \leq w_{max}$} 
            \State $\phi(v_i, v_j) \leftarrow \frac{\sum_{v_i \in e, v_j \in e}\frac{weight_e}{|e|-1}}{W_k[{v_i}] + W_{k+1}[cmap[{v_j}]]}$  ~~~// Cluster Score
            \If{$\phi(v_i, v_j) > \phi _{best}$} 
                $v_{best} \leftarrow v_j$
            \EndIf
        \EndIf
        \EndFor
        \If{$v_{best} == -1$}
        \State $W_{k+1}[v_i] \leftarrow W_k[v_i]$
        \State $visited[v_i] \leftarrow true$ 
        \Else
            \State $cmap[v_i] \leftarrow cmap[v_{best}]$
            \State $W_{k+1}[v_{best}] \leftarrow W_k[{v_i}] + W_{k+1}[cmap[v_{best}]]$
            \State $visited[v_i] \leftarrow true$; $visited[v_{best}] \leftarrow true$ 
            \State $N_{cluster} \leftarrow N_{cluster}-1$
        \EndIf
    \EndFor
    \If{$N_{cluster} == |V_{k-1}|$} 
        \State break
    \Else 
        \State $H_{k+1}(V_{k+1}, E_{k+1}, W_{k+1}) \leftarrow$ Build clustered hypergraph using $cmap$
    \EndIf
    \EndFor
    \State $H_c \leftarrow$ Clustered hypergraph generated at last iteration
    \State $H_{out} \leftarrow$ Best-fit bin packing on $H_c$
    \State \textbf{Return} $H_{out}$
\end{algorithmic}
\label{alg:clustering}
\end{algorithm}

\noindent
{\bf Clustered Hypergraph Creation.} 
For a given design, we first obtain a netlist hypergraph using 
OpenDB \cite{OpenDB}. We perform {\em cell width-regularized clustering}, where 
cells  $cells_{orig}$ (vertices) in the original netlist hypergraph are clustered 
such  that {\em clustered cell width}\footnote{Given vertex set $V$ with cell 
widths $W$, clustering vertices $v_i, v_j \in V$ yields a clustered cell with 
width $W[{v_i}]$ + $W[{v_j}]$.} does not exceed 
$w_{max}$, the maximum cell width in the library. 
The inputs to cell width-regularized clustering are (i) a hypergraph 
$H(V,E,W)$ with vertices $V$, hyperedges $E$ and cell widths $W$, 
(ii) the maximum cell width, $w_{max}$, 
and (iii) a limit on number of clustering iterations, $N_{iter}$.\footnote{In our
experiments, we set $N_{iter} = 20$. However, the cell width-regularized 
clustering is strongly constrained by $w_{max}$, and we observe on our testcases
that clustering stops after $\sim3$ iterations.}
The output is a clustered hypergraph ($H_{out})$. 
We use {\em First-Choice} ({\em FC}) clustering ~\cite{KarypisAKS99} 
and refer to our clustering method as {\em cell width-regularized clustering with FC}, or
{\em CWR-FC}. 

{\em CWR-FC} first sorts vertices in increasing order of cell widths 
(Line 4) and initializes {\em cluster assignments} (Line 6).
The cluster assignment $cmap$ is the mapping of vertices to clusters 
($V_k$ to $V_{k+1}$).  Clustered cell widths $W_{k+1}$ are initialized in Line 7.
Next, vertices are traversed in order to perform pairwise clustering;
note that only combinational cells are considered for clustering (Line 8). 
For each vertex $v_i$ that is traversed, we find its neighbors $v_j$ in the hypergraph
(Line 11). 
Each $v_j$ is considered only if it does not violate the $w_{max}$ limit (Line 14);
a cluster score $\phi(v_i,v_j)$ is calculated in Line 15.
In the cluster score, $weight_e$ is the weight of hyperedge $e$ and 
$W_k[v_i]$ is the width of vertex $v_i$. The numerator aims to cluster
vertices that are strongly connected (i.e., share many hyperedges) 
while the denominator promotes clusters of similar widths. 
If all neighboring vertices $v_j$ violate the threshold width constraint,
then no new clusters are formed (Lines 17-19). Otherwise,
the vertex with the highest cluster score is selected, and 
a new cluster is created (Lines 21-24). 
After all vertices are visited, we construct the clustered hypergraph and proceed 
with subsequent iterations (Line 28). 
If no further clustering is feasible, the process terminates (Line 26).   

Note that {\em CWR-FC} clusters vertices that are adjacent to each
other in the hypergraph. However, if all pairings of vertices selected 
for clustering violate the $w_{max}$ width constraint, the algorithm can stall 
(Line 25). 
To address this issue and improve the uniformity of cluster contents, we perform 
best-fit bin-packing \cite{JohnsonDS73} with bins having capacity $w_{max}$ 
(Line 30).\footnote{The choice of best-fit is motivated by its simplicity and 
intuitiveness. Best-fit also enjoys a better approximation ratio compared to 
first-fit or next-fit alternatives \cite{JohnsonDS73}.}
Finally, the output is the clustered hypergraph $H_{out}$.

\noindent
{\bf Clustered Netlist Creation.}
We convert the clustered hypergraph $H_{out}$ into Verilog using OpenDB. 
Then, to run P\&R we require a new LEF file that captures the cluster assignments
from cell width-regularized clustering. I.e., we require a new netlist over
the clusters, $cells_{clustered}$.

Figure~\ref{fig:toy_example}(a) provides a schematic view
of two clustered cells, NAND\_X1\_AND\_X1 and INV\_X1\_OR\_X1.
These correspond to two clusters of original cells: NAND\_X1 and AND\_X1, and 
INV\_X1 and OR\_X1. 
How the clustered cells are composed from original standard-cell layouts 
is shown in Figure~\ref{fig:toy_example}(b). 
In this case, a non-integer gear ratio between M1P (30nm) and CPP (45nm) forces
cells in $cells_{clustered}$ to be positioned at even CPP sites, to avoid M1 pin 
misalignment. 
In the first cluster, NAND\_X1 width (3CPP) is an odd number of CPPs,
necessitating addition of 1CPP padding between the two cells. 
In the second cluster, the total cell width is less than $w_{max}$,
so whitespace is included along with the clustered original cells.
We distribute whitespace uniformly, (i) at the sides of $cells_{clustered}$ 
and (ii) between consecutive cells in each cluster, as illustrated in
Figure~\ref{fig:toy_example}(b). 
During this whitespace allocation, we first allocate whitespace at junctions (between
consecutive original cells) where no extra padding has been previously allocated.

\vspace{-0.3cm}
\subsection{Performance of Clustered Cell Width-Regularization} 
\vspace{-0.1cm}

We now document advantages of our proposed clustered cell width-regularization, i.e.,
more realistic utilization in P\&R blocks, and realistic topological and wirelength 
characteristics of P\&R outcomes.

\begin{figure}[t]
\center
\includegraphics[width=0.93\columnwidth]{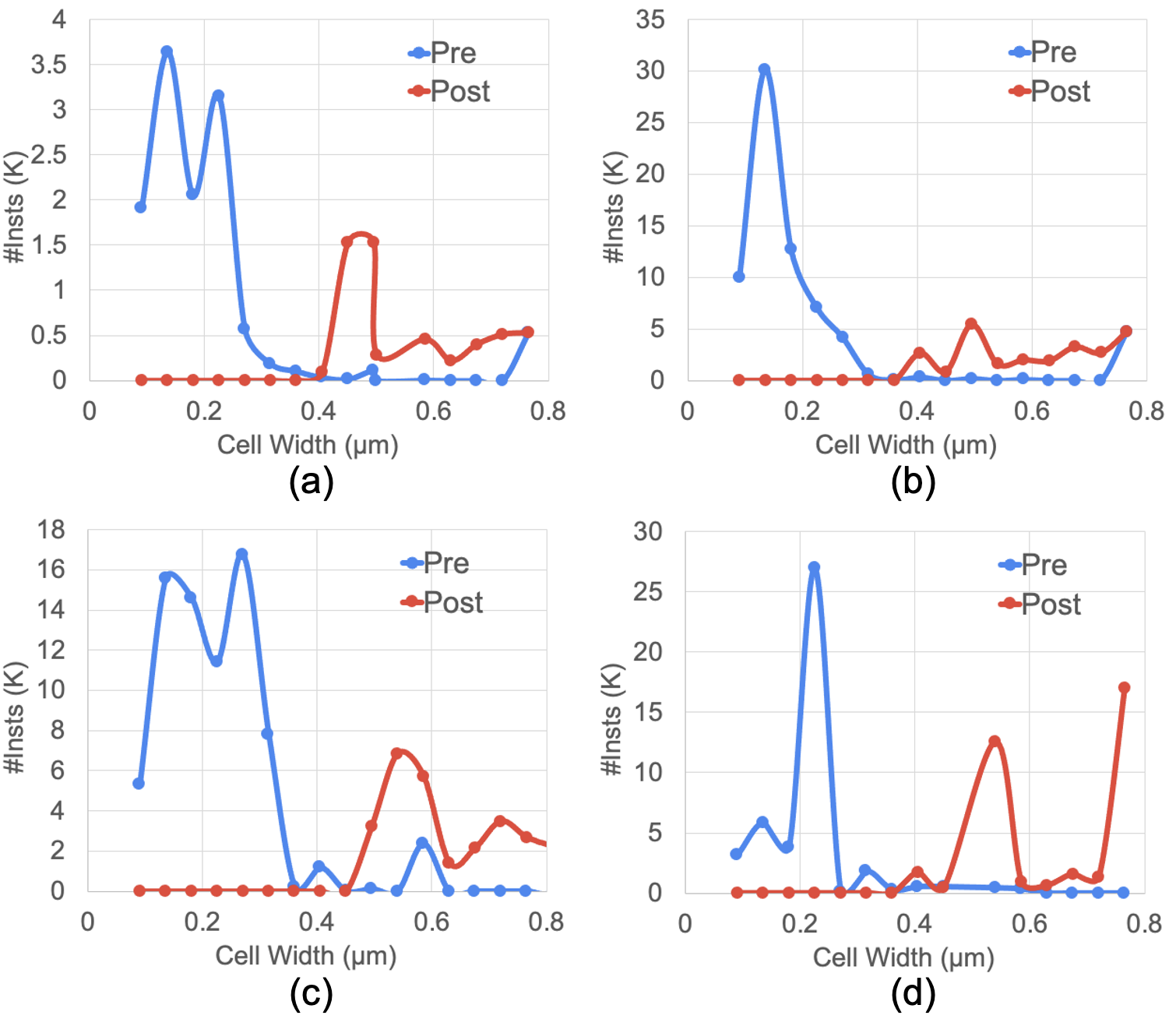}
\vspace{-0.3cm}
\caption{Cell width distributions pre-clustering (i.e., original netlist)
and post-clustering (i.e., by {\em CWR-FC}) for (a) AES, (b) JPEG, (c) LDPC and (d) VGA.} 
\vspace{-0.3cm}
\label{fig:width_comp}
\end{figure}

\noindent
{\bf Comparison to Previous Cell Width-Regularization.}
Figure~\ref{fig:width_comp} compares cell width distributions for instances in the 
clustered netlist and instances in the original netlist. 
The blue lines show the distribution of cell widths in the original netlist, 
where smaller cell widths predominate. 
The red lines indicate that {\em CWR-FC} increases the prevalence 
of cells with larger widths through creation of the merged $cells_{clustered}$.
The larger amount of actual cell widths in $cells_{clustered}$ leads to smaller
amounts of added whitespace needed to regularize cell widths.

As anticipated, {\em clustered} cell width-regularization significantly reduces whitespace
in the placed designs.
With {\em Lib2} and FSPDN for P\&R, placing cell width-regularized instances used 
in PROBE2.0 at 90\% density achieves actual utilizations of 0.21, 0.21 and 0.40 for AES, 
JPEG and VGA, respectively.  For LDPC, placing cell width-regularized instances at 30\% density
achieves actual utilization of 0.08.
By contrast, with clustered cell width-regularization, we achieve actual utilizations 
of 0.71, 0.74 and 0.71 for AES, JPEG and VGA, respectively.
For LDPC, we achieve actual utilization of 0.23.
In this way, our new methodology enables $K_{th}$ evaluation by iterated neighbor-swapping 
while preserving realistic placement utilizations.

\noindent
{\bf Topological and Wirelength Comparisons to Real Designs.}
We have confirmed additional similarities between between clustered cell 
width-regularized netlists and the original real designs.
Table~\ref{tab:compcwreg} compares
characteristics of our clustering-based cell width-regularized netlists and placements 
([C]), versus analogous characteristics of real netlists and placements ([A]). 
We also implement another plausible clustering methodology, which is to induce
clusters from a placement of the original design ([B]).
In [B], clusters from the placement are induced by (i) traversing combinational cells left-to-right
in each standard cell row, and (ii) clustering maximal contiguous sets of cells
without exceeding $W_{max}$.

We run P\&R using {\em Lib2} and $P_{FS}$ for PDN, maintaining the same core 
area and utilization.  Clustering decreases the number of instances and 
average fanouts for [B] and [C], relative to [A].
However, wirelengths exhibit no significant changes. 
The similarities between [A], [B] and [C] suggest that our 
{\em CWR-FC} methodology can preserve netlist properties relevant to
P\&R outcomes, with more realistic placement utilizations.

\begin{table}[t]
\vspace{-0.2cm}
\caption{Comparison of the width-regularized clustered netlist produced by {\em CWR-FC} ([C]) 
with the original flat netlist ([A]) and a width-regularized clustered netlist induced
from a placement of the flat netlist ([B]).}
\vspace{-0.4cm}
\scriptsize
\label{tab:compcwreg}
\begin{center}
\begin{tabular}{|c|c||c|c|c|c|c|}
\hline
{\bf Stage} & {\bf Design}  & {\bf \#Insts} & {\bf Area ($um^2$)} & {\bf Util} & {\bf WL ($um$)} & {\bf Avg. FO}\\\hline
\multirow{4}{*}{[A]} & AES & 12318 & 426.254 & 0.83 & 30849 & 2.32\\\cline{2-7}
 & JPEG & 70031 & 2781.981 & 0.73 & 112605 & 2.15\\\cline{2-7}
 & LDPC & 77379 & 6250.563 & 0.43 & 567630 & 1.85\\\cline{2-7}
 & VGA & 60921 & 4238.205 & 0.76 & 208845 & 2.71\\\hline
\multirow{4}{*}{[B]} & AES & 4275 & 426.254 & 0.83 & 32632 & 1.96\\\cline{2-7}
 & JPEG & 23281 & 2781.981 & 0.73 & 111241 & 1.86\\\cline{2-7}
 & LDPC & 42383 & 6250.563 & 0.43 & 585923 & 1.43 \\\cline{2-7}
 & VGA & 40084 & 4238.205 & 0.76 & 189612 & 2.14 \\\hline
 \multirow{4}{*}{[C]} & AES & 4661 & 426.254 & 0.83 & 32679 & 2.08\\\cline{2-7}
 & JPEG & 25961 & 2781.981 & 0.73 & 143693 & 1.76\\\cline{2-7}
 & LDPC & 30636 & 6250.563 & 0.43 & 637417 & 1.29 \\\cline{2-7}
 & VGA & 32768 & 4238.205 & 0.76 & 220915 & 2.02 \\\hline
\end{tabular}
\end{center}
\vspace{-0.2cm}
\end{table}

\vspace{-0.3cm}
\section{Experimental Setup and Results}
\label{sec:expt}
\vspace{-0.1cm}

We have extensively studied the design-technology pathfinding 
capability of the PROBE3.0 framework using the PROBE3.0 technology. 
In this section, we report three main experiments.
Expts 1 and 2 show PROBE3.0's capability to assess PPAC trends and tradeoffs,
using real and artificial designs respectively.
Expt 3 performs assessments of routability and achievable utilization.

In Expts 1 and 2, we analyze four tradeoffs. 
(i) We present {\em Performance-Power} 
plots that quantify tradeoffs between performance (maximum frequency) and power.
(ii) We present {\em Performance-Area} plots to quantify the tradeoffs between
performance and area.
(iii) To address PP aspects, we use the {\em Energy-Delay Product} (EDP) 
~\cite{LiebmannCCC20} as a single metric for power and performance.
{\em EDP-Area} plots depict tradeoffs between performance/power and area.
(iv) We present {\em IR drop-Area} plots to demonstrate tradeoffs between 
IR drop and area. We also compare results obtained using artificial 
designs with those obtained using real designs.
Expt 3 assesses routability and achievable utilization
using our clustering-based cell width-regularized placements.

\vspace{-0.3cm}
\subsection{Experimental Setup}
\label{subsec:exptsetup}
\vspace{-0.1cm}

Based on the definition of technology and 
design parameters in~\cite{ChengKKK22},  we define ten technology 
parameters and eight design parameters as the input parameters for
the PROBE3.0 framework.
Table~\ref{tab:param} describes the definitions of these parameters 
and the options used in our experiments. 
Also, we use commercial tools for PDK generation, 
logic synthesis, P\&R, and IR drop analysis. We use open-source 
tools for GDT-to-GDS translation \cite{gdt2gds} and SMT solver \cite{Z3}.
Table~\ref{tab:tool} summarizes the tools and versions that we use in 
our experiments.

\begin{table*}[t]
\vspace{-0.3cm}
\caption{Technology and design parameters in our experiments.}
\vspace{-0.3cm}
\scriptsize
\label{tab:param}
\begin{center}
\begin{tabular}{|c|c|p{10cm}|l|}
\hline
{\bf Type}                  & {\bf Parameter}  & {\bf Description} & {\bf Option}\\\hline
\multirow{15}{*}{Technology} & $Fin$ & The number of fins for devices of standard cells. & 2, 3   \\\cline{2-4}
                            & $CPP$ & Contacted poly pitch for standard cells in $nm$. & 45 \\\cline{2-4}
                            & $M0P$ & M0 (horizontal) layer pitch in $nm$. & 24 \\\cline{2-4}       
                            & $M1P$ & M1 (vertical) layer pitch in $nm$. & 30 \\\cline{2-4} 
                            & $M2P$ & M2 (horizontal) layer pitch in $nm$. & 24 \\\cline{2-4} 
                            & $RT$ & The number of available M0 routing tracks in
standard cells. & 4, 5  \\\cline{2-4}
                            & $PGpin$ & Power/ground pin layer for standard cells. & $BPR$, $M0$ \\\cline{2-4}
                            & \multirow{3}{*}{$CH$} & Cell height of standard cells, expressed as a multiple
of $M0P$. For example, when the cell height in $nm$ is
$120nm$ and $M0P$ is $24nm$, the cell height ($CH$) is
5. The cell height value is calculated as $RT + 2$ for $M0$
$PGpin$ and $RT + 1$ for $BPR$ $PGpin$. & \multirow{3}{*}{5, 6, 7} \\\cline{2-4}
                            & $MPO$ & The number of minimum pin openings (access
points). & 2 \\\cline{2-4}
                            & \multirow{4}{*}{$DR$} & Design rules. We define the same grid-based 
    design rules, minimum area rule ({\em DR-MAR}), 
    end-of-line spacing rule ({\em DR-EOL}) and via spacing rule ({\em DR-VR}) 
    as~\cite{ChengKKK22}. We use the {\em EUV-tight} ($ET$) 
    design rule set, which includes {\em DR-MAR} $= 1$, {\em DR-EOL} $= 2$ and {\em DR-VR} $= 1$. & \multirow{4}{*}{{\em EUV-Tight}}\\\hline
\multirow{11}{*}{Design}     & \multirow{2}{*}{$BEOL$} & Metal stack options. We define 14M metal option which contains 
14 metal layers (M0 to M13). 
We define 1.2X, 2.6X, 3.2X and 30X layer pitches based on 24nm as the 1X pitch. & \multirow{2}{*}{14M} \\\cline{2-4}
                            & $PDN$ & Power delivery network options. & $P_{FS}$, $P_{FB}$, $P_{BS}$, $P_{BB}$ \\\cline{2-4}
                            & $I_{pitch}$ & Power tap cell pitch in CPP. & 24, 32, 48, 96, 128 \\\cline{2-4}
                            & $I_{scheme}$ & Power tap cell insertion scheme. & {\em Column}, {\em Staggered} \\\cline{2-4}
                            & $Tool$ & Commercial P\&R tools. & {\em Synopsys IC Compiler II} \\\cline{2-4}
                            & $Util$ & Initial placement utilization. & 0.70 to 0.94 with a 0.02 step size\\\cline{2-4}
                            & \multirow{2}{*}{$Design$} & Designs studied in our experiments. We conduct experiments with 
    four open-source designs from OpenCores~\cite{OpenCores} and artificial netlists 
    generated by ANG with our ML-based parameter tuning.  & \multirow{2}{*}{AES, JPEG} \\\cline{2-4}
                            & \multirow{2}{*}{$Clkp$} & Target clock period for logic synthesis and P\&R. 
    We define target clock periods 
    that reflect maximum achievable frequencies of the designs. & \multirow{2}{*}{0.12 to 0.24ns with a 0.02ns step size}\\\hline
\end{tabular}
\end{center}
\vspace{-0.5cm}
\end{table*}

\begin{table}[t]
\vspace{-0.3cm}
\caption{Tools and versions in our experiments.}
\vspace{-0.3cm}
\scriptsize
\label{tab:tool}
\begin{center}
\begin{tabular}{|c|c|c|c|}
\hline
{\bf Purpose} & {\bf Tool} & {\bf Version} & {\bf Ref.} \\\hline
Format Conversion & {\em GDT-to-GDS translator} & 4.0.4 & \cite{gdt2gds}\\\hline
IR Drop Analysis & {\em Cadence Voltus} & 19.1 & \cite{Voltus}\\\hline
Library Characterization & {\em Cadence Liberate} & 16.1 & \cite{Liberate}\\\hline
\multirow{2}{*}{Logic Synthesis} & {\em Cadence Genus} & 21.1 & \cite{Genus}\\\cline{2-4}
& {\em Synopsys Design Compiler} & R-2020.09 & \cite{DC}\\\hline
LVS & {\em Siemens Calibre} & 2017.4\_19 & \cite{Calibre}\\\hline
P\&R & {\em Synopsys IC Compiler-II} & R-2020.09 & \cite{ICC2}\\\hline
\multirow{3}{*}{PEX} & {\em Cadence QRC Extraction} & 19.1 & \cite{QRC}\\\cline{2-4}
 & {\em Synopsys StarRC} & O-2018.06 & \cite{StarRC}\\\cline{2-4}
 & {\em Siemens Calibre} & 2017.4\_19 & \cite{Calibre}\\\hline
SMT solver & {\em Z3} & 4.8.5 & \cite{DeMouraB08}\cite{Z3}\\\hline
\end{tabular}
\end{center}
\vspace{-0.6cm}
\end{table}

\noindent
{\bf Criteria for Valid Result.}
In our experiments, for given $Design$, $PDN$ and technology parameters, 
we perform logic synthesis, P\&R and IR drop analysis with multiple sets of parameters including $I_{pitch}$, $I_{scheme}$, $Util$ and $Clkp$.
We use 24, 32, 48, 96 and 128 $CPP$ for $I_{pitch}$, and 
{\em Column} and {\em Staggered} for $I_{scheme}$.
For $Util$, we use values ranging from 0.70 to 0.94 with a step size 
of 0.02, and for $Clkp$, we use values ranging from 0.12 to 0.24ns with 
a step size of 0.02ns.
Importantly, after the implementation and the analysis steps,
we filter out results that are deemed invalid -- in that they
are likely to fail signoff criteria even with additional human 
engineering efforts. 

To be precise, a ``valid'' result must satisfy three conditions: 
(i) the worst negative slack is larger than -50ps;
(ii) the number of post-route DRCs is less than 500; and
(iii) the 99.7 percentile of the effective instance voltage is 
greater than 80\% of the operating voltage ($V_{op}$).
To assess (iii), we use a commercial IR drop analysis 
tool~\cite{Voltus} to measure vectorless dynamic IR drop,
and calculate the effective instance voltage as 
$V_{op} - V_{drop}$ per each instance, where $V_{op}$ is an operating 
voltage (0.7V) and $V_{drop}$ is the worst voltage drop per instance.
We take the 99.7 percentile of effective instance voltage
as representative of IR drop for the post-P\&R result, as it is
within three standard deviations from the mean per the
empirical rule~\cite{EmpiricalRule}.

\vspace{-0.3cm}
\subsection{Expt 1 (PPAC Exploration with Real Designs)}
\vspace{-0.1cm}

\noindent
{\bf Performance versus Power.}
We first present PPAC explorations that show tradeoffs between 
performance and power. (We assume that area is proportional 
to cost, since chip area is closely related to cost.)
In this study, we show results for JPEG with four standard-cell 
libraries ({\em Lib1-4}).
Also, we use four PDN structures, $P_{FS}$, $P_{FB}$, $P_{BS}$ and 
$P_{BB}$, and measure improvements due to scaling boosters
relative to the traditional frontside PDN ($P_{FS}$). 

\begin{figure*}[htb!]
\center
\includegraphics[width=2.0\columnwidth]{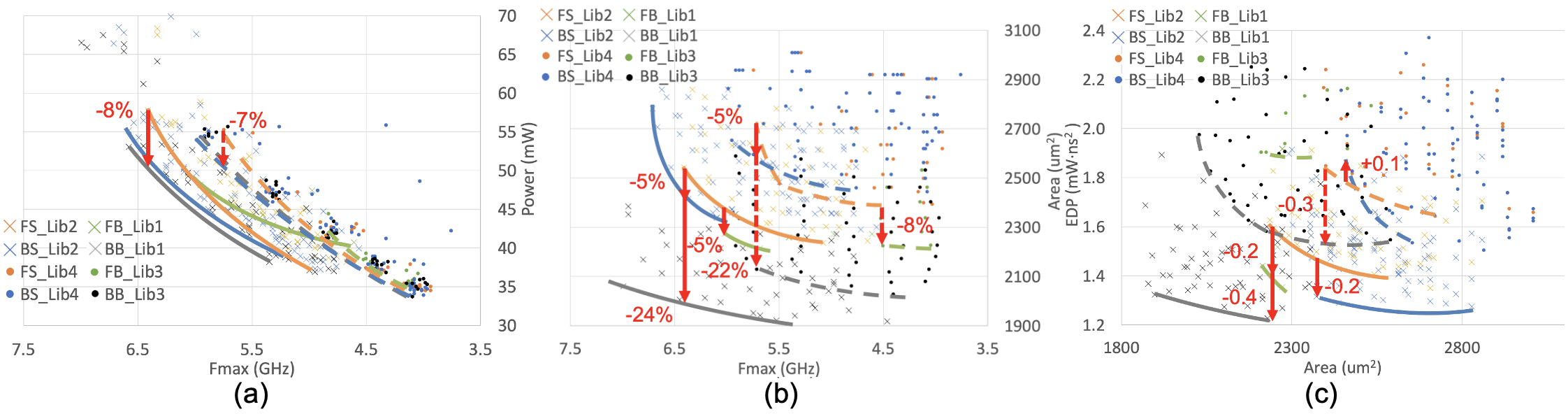}
\vspace{-0.4cm}
\caption{PPAC tradeoffs for JPEG with
four standard-cell libraries ({\em Lib1}, {\em Lib2}, {\em Lib3} and {\em Lib4}). 
We compare four PDN structures 
in terms of performance and area and measure improvements relative
to traditional frontside PDN ($P_{FS}$), in order to show the benefits 
of the scaling boosters (BSPDN and BPR). 
We draw plots for (a) {\em Performance-Power}, (b) {\em Performance-Area},
and (c) {\em Energy-Delay Product-Area}.}
\vspace{-0.4cm}
\label{fig:ppac_jpeg}
\end{figure*}

Figure~\ref{fig:ppac_jpeg}(a) gives {\em Performance-Power} plots that
show tradeoffs between performance and power for JPEG, and improvements 
from the traditional FSPDN.
We calculate the maximum achievable frequency ($f_{max}$) as 
$1/(Clkp-WNS)$ where $Clkp$ is the target clock period and $WNS$ is the 
worst negative slack. 
Also, we add up leakage and dynamic power to obtain the total power.
To measure the improvement from $P_{FS}$, we compare the second-largest
value (on the x-axis) attained with each PDN configuration.
From the result, we make two main observations.
(i) Power consumption with $P_{BS}$ and $P_{BB}$ decreases by 7 to 8\%,
compared to $P_{FS}$ with the same performance.
(ii) Power consumption with $P_{FB}$ is similar to $P_{FS}$, with 
the same performance.
We observe power reductions from use of scaling boosters, BSPDN and BPR.
However, use of BPR without BSPDN does not reduce power consumption.

\noindent
{\bf Performance versus Area.}
Performance-area tradeoffs 
for JPEG are shown in Figure~\ref{fig:ppac_jpeg}(b).
We make two main observations.
(i) Area with $P_{FB}$, $P_{BS}$ and $P_{BB}$ decreases 
by up to 8\%, 5\% and 24\%, respectively, as
compared to $P_{FS}$, while maintaining the same level of performance.
(ii) We find that use of scaling boosters results in area reductions 
across all four standard-cell libraries.
The area reduction results obtained using the PROBE3.0 framework
are consistent with previous industry 
works~\cite{HossenCVB20}\cite{Ryckaert19}\cite{BSBPRAM21}\cite{BSBPRarticle22}, 
which show that use of BSPDN and BPR techniques can result in area 
reductions of 25\% to 30\%.

\noindent
{\bf Energy-Delay Product (EDP) versus Area.}
Given the tradeoffs among PPAC criteria, a simpler metric is useful to 
comprehend multiple aspects simultaneously.
The {\em Energy-Delay Product} (EDP) is adopted by, e.g., 
\cite{LiebmannCCC20} as a single-value metric that captures both
power efficiency and maximum achievable frequency (performance).
EDP is calculated as $P \times {f_{max}}^2$, where $P$ denotes power consumption
and $f_{max}$ denotes maximum achievable frequency.
Lower EDP means more energy-efficient operations for the chip.
Since we address power, performance and area (cost), we draw 
{\em EDP-Area} plots to show PPAC tradeoffs of various PDN structures.
We again use four standard-cell libraries ({\em Lib1-4}).

From Figure~\ref{fig:ppac_jpeg}(c), we derive four key observations.
(i) For $4RT$ ({\em Lib1} and {\em Lib2}), EDP with $P_{FB}$, $P_{BS}$ and $P_{BB}$ 
decreases by 0.2, 0.2 and 0.4 $mW \cdot ns^2$, respectively, 
compared to $P_{FS}$ with the same area.
(ii) For $5RT$ ({\em Lib3} and {\em Lib4}), EDP with $P_{BB}$ decreases 
by 0.3 $mW \cdot ns^2$, compared to $P_{FS}$ with the same area.
(iii) For $5RT$, EDP with $P_{FB}$ shows no improvements, 
and EDP with $P_{BS}$ increases by 0.1 $mW \cdot ns^2$, as compared 
to $P_{FS}$ with the same area.
(iv) Use of $P_{BB}$ better optimizes area than other PDN structures 
with the same EDP.

\begin{figure}[t]
\center
\includegraphics[width=1.0\columnwidth]{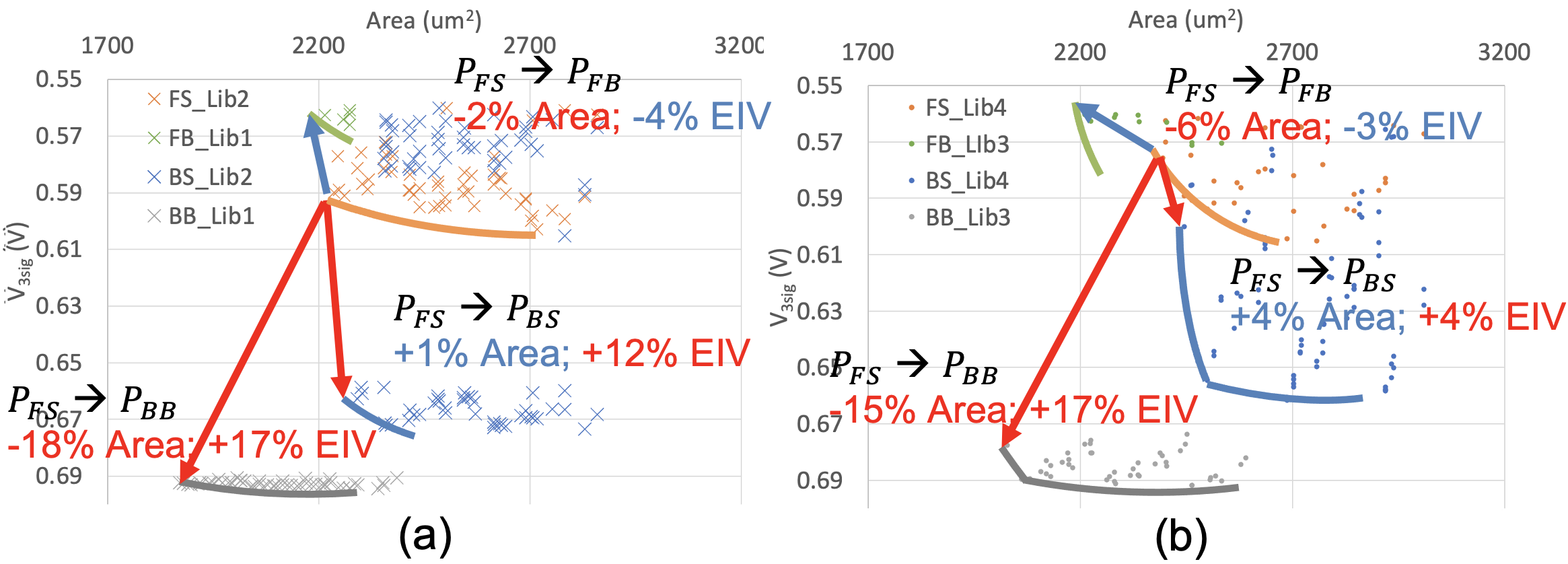}
\vspace{-0.4cm}
\caption{{\em IR drop-Area} plots for JPEG 
with four standard-cell libraries ({\em Lib1}, {\em Lib2}, {\em Lib3} and {\em Lib4}). 
(a) JPEG with $4RT$ ({\em Lib1/2}),
and (b) JPEG with $5RT$ ({\em Lib3/4}).}
\vspace{-0.3cm}
\label{fig:ir_area}
\end{figure}

\noindent
{\bf Supply Voltage (IR) Drop versus Area.}
With recent advanced technologies and designs, denser PDN structures 
are required due to large resistance seen in tight-pitch BEOL metal 
layers. The denser PDN structures bring added routability challenges 
which critically impact area density.
In light of this, we measure IR drop and area from valid 
runs, and plot {\em IR drop-Area} tradeoffs 
in Figure~\ref{fig:ir_area}.
In the plots, we compare the points with the minimum area for 
each PDN configuration in terms of area and 99.7 percentile 
(three-sigma) of effective instance voltage (EIV).
Note that larger effective instance voltage means better IR drop mitigation.
Figures~\ref{fig:ir_area}(a) and (b) show {\em IR drop-Area} 
tradeoffs for JPEG with $4RT$ ({\em Lib1} and {\em Lib2}) and $5RT$ ({\em Lib3} and {\em Lib4}), respectively.
From the results, we make four main observations. 
(i) Area with $P_{FB}$ decreases by 2 to 6\% compared to $P_{FS}$, while the effective instance voltage (EIV) increases by 3 to 4\%.
(ii) Area with $P_{BS}$ increases by 1 to 4\% compared to 
$P_{FS}$, while EIV decreases by 4 to 12\%.
(iii) Area with $P_{BB}$ decreases by 15 to 18\% compared to 
$P_{FS}$, while EIV decreases by 17\%.
(iv) We observe that there is IR drop mitigation from use of backside 
PDN, while use of BPR ($P_{FB}$) worsens IR drop. This implies that 
more power tap cells will need to be inserted to mitigate IR drop. 
However, the area overhead of power tap cells will degrade the 
IR drop quality achieved by use of BPR.

\vspace{-0.3cm}
\subsection{Expt 2 (PPAC Exploration with Artificial Design)}
\label{subsec:angppac}
\vspace{-0.1cm}

Our second main experiment uses the {\em artificial} JPEG design 
generated by ANG using our ML-based parameter tuning.
We conduct the same studies as in Expt 1 and analyze the results. 

\begin{figure*}[htb!]
\center
\includegraphics[width=2.0\columnwidth]{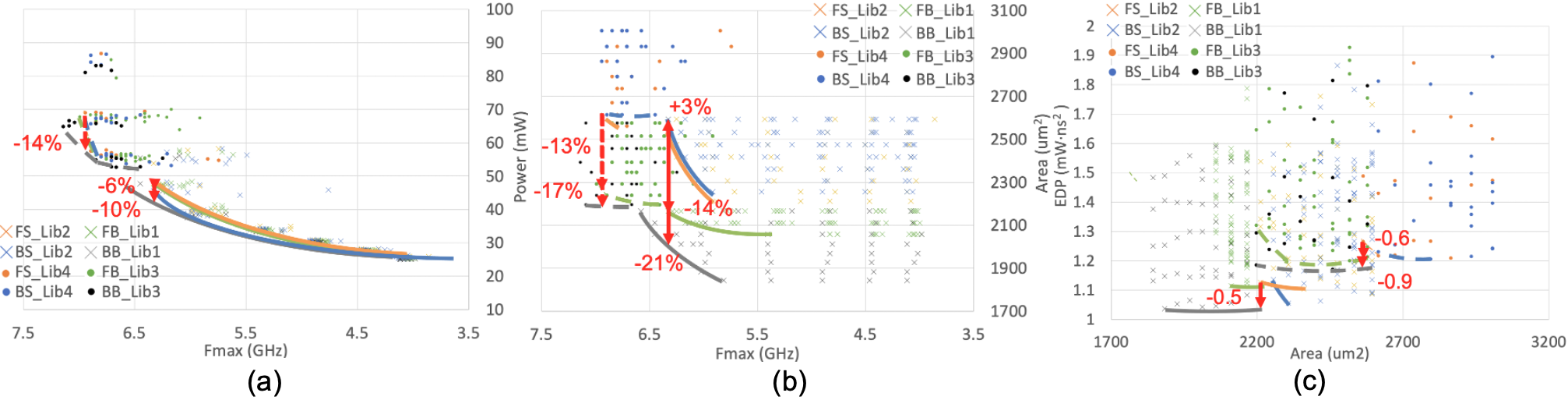}
\vspace{-0.3cm}
\caption{PPAC tradeoffs for ``artificial'' JPEG with
four standard-cell libraries ({\em Lib1}, {\em Lib2}, {\em Lib3} and {\em Lib4}). 
Shown: (a) {\em Performance-Power}, (b) {\em Performance-Area},
and (c) {\em Energy-Delay Product-Area}.}
\vspace{-0.3cm}
\label{fig:artppa}
\end{figure*}

\noindent
{\bf Performance versus Power.}
Figure \ref{fig:artppa}(a) shows the tradeoffs between performance and 
power with the artificial JPEG design.
From the result, we make three main observations. (i) Power consumption
with $P_{BS}$ and $P_{BB}$ decreases by 6 to 14\%, compared to $P_{FS}$ with
the same performance. (ii) Power consumption with $P_{FB}$ is
similar to $P_{FS}$ with the same performance. 
(iii) Results with the artificial JPEG show up to 7\% differences,
but with similar trends, compared to the results obtained with the 
real JPEG design.

\noindent
{\bf Performance versus Area.}
Figure \ref{fig:artppa}(b) shows tradeoffs between performance and 
area with the artificial JPEG design.
We make three main observations.
(i) Area with $P_{FB}$ and $P_{BB}$ decreases up to 14\% and 21\%
compared to $P_{FS}$ with the same performance.
(ii) Area with $P_{BS}$ increases by 0\% to 3\%
compared to $P_{FS}$ with the same performance. This area penalty is 
caused by power tap cell insertion for $P_{BS}$.
(iii) We observe that the results with the artificial JPEG show up to 
9\% differences, but with similar trends, compared to the results obtained
with the real JPEG design.
However, area for $P_{BS}$ shows opposite trends to what we observe
with the real design, although the discrepancy is not too large.

\noindent
{\bf Energy-Delay Product (EDP) versus Area.}
From Figure~\ref{fig:artppa}(c), we make three main observations.
(i) For $4RT$ ({\em Lib1} and {\em Lib2}), EDP with $P_{BB}$ 
decreases by 0.5 $mW \cdot ns^2$, compared to $P_{FS}$ with the same area.
However, EDP with $P_{FB}$ and $P_{BS}$ shows no improvements.
(ii) For $5RT$ ({\em Lib3} and {\em Lib4}), EDP with $P_{FB}$ and $P_{BB}$ decreases 
by 0.6 and 0.9 $mW \cdot ns^2$, compared to $P_{FS}$ with the same area.
However, EDP with $P_{BS}$ shows no improvements.
(iii) We observe that results with the artificial JPEG show
similar trends as results obtained with the real JPEG design.

\begin{figure}[t]
\center
\includegraphics[width=1.0\columnwidth]{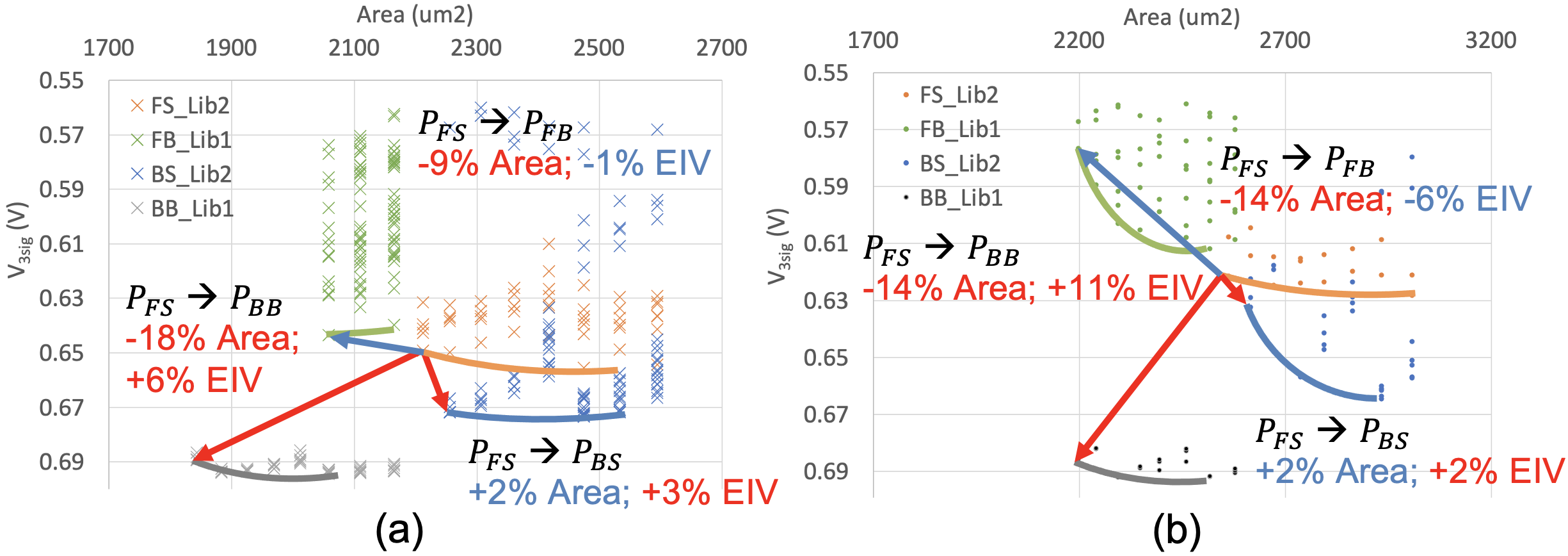}
\vspace{-0.3cm}
\caption{{\em IR drop-Area} plots for ``artificial'' JPEG 
with four standard-cell libraries ({\em Lib1}, {\em Lib2}, {\em Lib3} 
and {\em Lib4}). 
(a) JPEG with $4RT$ ({\em Lib1/2}), and (b) JPEG with $5RT$ ({\em Lib3/4}).}
\vspace{-0.4cm}
\label{fig:art_ir_area}
\end{figure}

\noindent
{\bf Supply Voltage (IR) Drop versus Area.}
Figures~\ref{fig:art_ir_area}(a) and (b) show tradeoffs 
between IR drop and area for the artificial JPEG design with 
$4RT$ ({\em Lib1} and {\em Lib2}) and $5RT$ ({\em Lib3} and {\em Lib4}), respectively. 
We make four main observations.
(i) Area with $P_{FB}$ decreases by 9 to 14\%, 
compared to $P_{FS}$, while the effective instance voltage (EIV) increases by 1 to 6\%.
(ii) Area with $P_{BS}$ increases by 2\%, compared to $P_{FS}$, while EIV decreases by 2 to 3\%.
(iii) Area with $P_{BB}$ decreases by 14 to 18\%, compared to $P_{FS}$, while EIV decreases by 6 to 11\%.
(iv) We observe that results with the artificial JPEG show similar trends 
as results obtained with the real JPEG design, and that discrepancies
are reasonably small.

\vspace{-0.3cm}
\subsection{Expt 3 (Routability Assessment and Achievable Utilization)}
\label{subsec:kth}
\vspace{-0.1cm}

Our third main experiment measures $K_{th}$ using our
{\em clustering-based cell width-regularized placements} (Section \ref{sec:enhancedcw}),
and explores the relationship between $K_{th}$ and achievable utilization. 
We note that the previous work of ~\cite{ChengKKK22} introduced
{\em Achievable Utilization} as the maximum utilization for which the 
number of DRCs is less than a predefined threshold of 500 DRCs.
Here, we include all three criteria for a valid result (Section \ref{subsec:exptsetup}),
and define {\em Achievable Utilization} as the maximum utilization among 
all valid runs seen. 

\begin{figure}[t]
\center
\includegraphics[width=1.0\columnwidth]{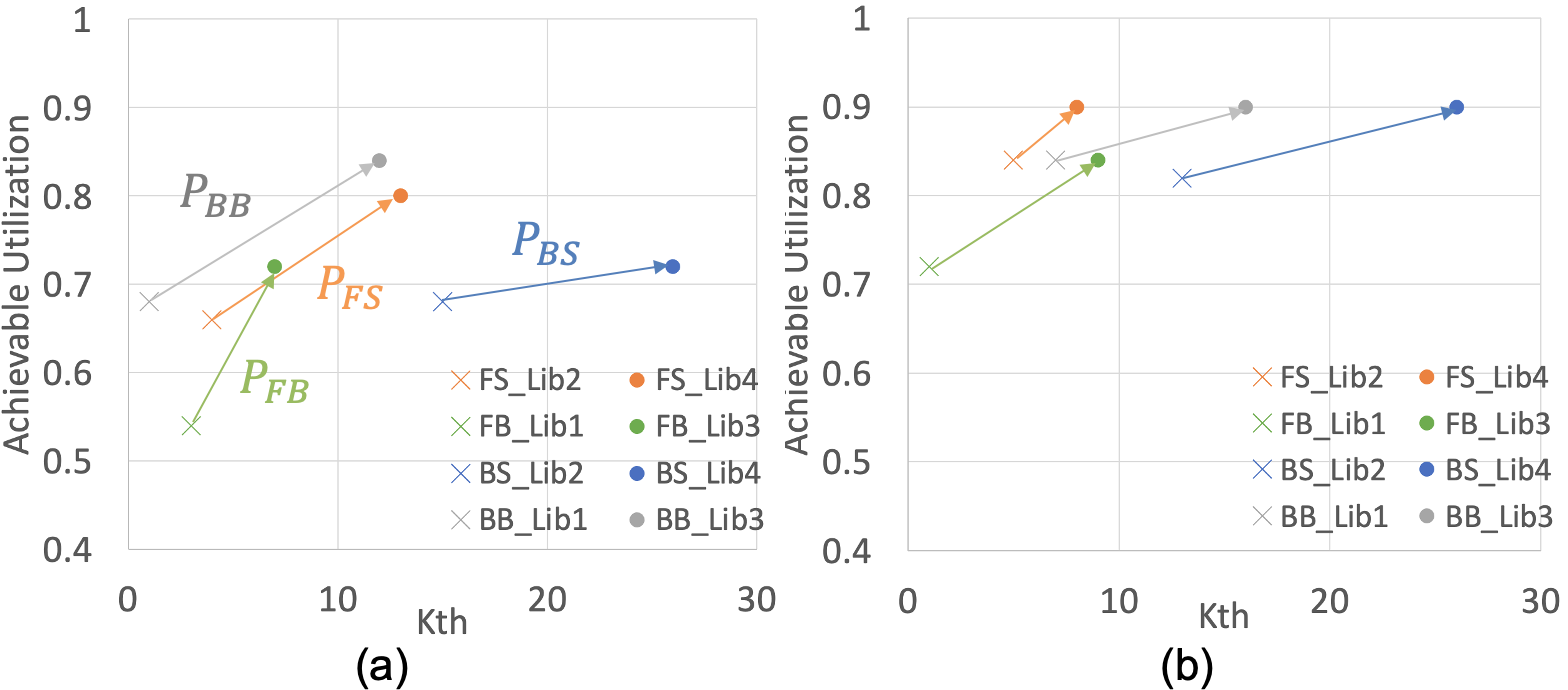}
\vspace{-0.5cm}
\caption{$K_{th}$ and achievable utilization for (a) AES
and (b) JPEG, with various libraries and power delivery methodologies.} 
\vspace{-0.5cm}
\label{fig:kth}
\end{figure}

Figure~\ref{fig:kth} shows experimental results for $K_{th}$ and 
achievable utilization. We conduct our experiment with artificial JPEG 
and four cell width-regularized libraries ({\em Lib1-4}).
From the plots, we make two observations.
(i) We compare the results with $2Fin$/$4RT$ standard-cell 
libraries ({\em Lib1/2}) to those with $3Fin$/$5RT$ standard-cell libraries 
({\em Lib3/4}). The data show that a larger number of M0 routing tracks 
brings better routability. 
(ii) Compared to $P_{FS}$, $P_{FB}$ and $P_{BB}$, the plots for 
$P_{BS}$ are skewed to the right for each design,
showing better routability than the other PDN configurations.
We observe that the routability improvement of $P_{BS}$ comes from 
regularly-placed power tap cells: the power tap cell placement 
eases routing congestion caused by high cell and/or pin density. 

Finally, we compare the $K_{th}$ results obtained with the previous
{\em cell width-regularized placements} used in the PROBE2.0 work ([A]) 
and  {\em clustering-based cell width-regularized placements} obtained
using the {\em CWR-FC} algorithm of Section \ref{subsec:Clustered} ([C]). 
We perform routability assessments as summarized in Table~\ref{tab:kthcomp}.
We rank-order $K_{th}$ across the eight combinations of four $PDN$
and two $RT$ with the JPEG design.
The main observation from this comparison is that the ordering of 
enablements based on $K_{th}$ is the same for both placements, even as
the area utilization of the {\em clustering-based cell width-regularized 
placements} is closer to the initial utilization (0.6).
We conclude that our {\em clustering-based cell width-regularized 
placement} methodology successfully provides more realistic placements 
without disrupting the $K_{th}$-based rank-ordering of enablements.
Moreover, the generally smaller $K_{th}$ values seen in the rightmost 
two columns of Table~\ref{tab:kthcomp} imply fewer P\&R trials needed to
evaluate the $K_{th}$ metric.

\begin{table}[t]
\vspace{-0.2cm}
\caption{$K_{th}$ comparison for the JPEG design with {\em cell width-regularized placements} ([A])
and {\em clustering-based cell width-regularized placements} ([C]).
$Util$ denotes real utilization with 0.6 initial utilization.
}
\vspace{-0.3cm}
\scriptsize
\label{tab:kthcomp}
\begin{center}
\begin{tabular}{|c|c|c|c||c|c||c|c|}
\hline
\multirow{2}{*}{{\bf Rank}} & \multirow{2}{*}{$PDN$} & \multirow{2}{*}{$RT$} & \multirow{2}{*}{{\bf Library}} & \multicolumn{2}{c||}{[A]} & \multicolumn{2}{c|}{[C]} \\\cline{5-8}
 & & & & {\bf $K_{th}$} & $Util$ & {\bf $K_{th}$} & $Util$ \\\hline
1 & $P_{FB}$ & 4 & {\em Lib1} & 6 & 0.14 & 3 & 0.50 \\\hline
2 & $P_{FS}$ & 4 & {\em Lib2} & 9 & 0.14 & 5 & 0.49  \\\hline
3 & $P_{BB}$ & 4 & {\em Lib1} & 12 & 0.14 & 7 & 0.50 \\\hline
4 & $P_{FS}$ & 5 & {\em Lib4} & 15 & 0.14 & 8 & 0.50 \\\hline
5 & $P_{FB}$ & 5 & {\em Lib3} & 16 & 0.14 & 9 & 0.50 \\\hline
6 & $P_{BS}$ & 4 & {\em Lib2} & 17 & 0.14 & 13 & 0.49 \\\hline
7 & $P_{BB}$ & 5 & {\em Lib3} & 18 & 0.14 & 16 & 0.50 \\\hline
8 & $P_{BS}$ & 5 & {\em Lib4} & 23 & 0.14 & 26 & 0.50 \\\hline
\end{tabular}
\vspace{-0.3cm}
\end{center}
\end{table}

\vspace{-0.3cm}
\section{Conclusion}
\label{sec:conc}
%\vspace{-0.1cm}

We have presented PROBE3.0, a systematic and configurable 
framework for ``full-stack'' PPAC exploration and pathfinding in 
advanced technology nodes. We introduce automated PDK and 
standard-cell library generation, along with enablement of 
scaling boosters in a predictive 3nm technology.
Our work is permissively open-sourced in GitHub \cite{PROBE3.0},
and includes open-sourceable PDKs and EDA tool scripts that 
incorporate power and performance considerations into the framework. 

We employ artificial netlist generation with a machine 
learning-based parameter tuning to mimic properties of arbitrary 
real designs. Along with a new {\em CWR-FC} clustering-based 
width-regularized netlist and placement methodology, this enables
PPAC exploration of a much wider space of technology, design enablement, 
and design options.
From our experiments, we find that the use of backside power delivery network 
(BSPDN) and buried power rails (BPR) can lead to up to 8\% reduction
in power consumption and up to 24\% reduction in area
using our predictive 3nm technology. 
These results from PROBE3.0 closely match previous works 
~\cite{HossenCVB20}\cite{Ryckaert19}\cite{BSBPRAM21}\cite{BSBPRarticle22} 
which estimated 25\% to 30\% area reduction from use of BSPDN and BPR.

Ongoing and future directions include the following.
(i) Improving the software architecture of PROBE3.0 will make it more
accessible and flexible for users to pursue their own PPAC explorations. 
Supporting the addition of user-defined variables can help 
capture and study variant technology and design assumptions.
(ii) To improve robustness of the framework, and its usefulness as
a ``proxy'' in real-world advanced technology development and DTCO,
improved device models, parasitic extraction models, signoff corner
definitions, relevant design examples, etc. will be beneficial.
It will also be necessary to add generation of DRC rule decks for 
commercial tools.
(iii) While our scripts for commercial tools are shared publicly in
our GitHub repository, using these tools still requires valid licenses. 
Incorporation of open-source tools into the PROBE3.0 framework 
can potentially lead to highly-scaled deployments, shorter 
turnaround times, and improved utility to a broader audience.

\vspace{-0.3cm}
\section*{Acknowledgments}
\vspace{-0.1cm}

We thank Dr. Mustafa Badaroglu at Qualcomm, 
Dr. Gi-Joon Nam at IBM, Dr. S. C. Song at Google 
and Prof. Taigon Song at Kyungpook National University
for valuable discussions.

\end{document}